\documentclass[aps,prd,twocolumn,nofootinbib,superscriptaddress,showkeys]{revtex4}
\usepackage{amssymb,amsmath,graphicx,enumerate}
\usepackage{subfigure}
\usepackage{dcolumn,booktabs}
\usepackage{longtable}
\usepackage[dvipdfm,colorlinks=true,citecolor=blue,pdfstartview=FitH]{hyperref}

\usepackage{color,framed} 
\definecolor{shadecolor}{rgb}{0.3,0.3,1} 

\def\bea{\begin{equation}}
\def\eea{\end{equation}}
\newcommand{\rt}{Regge trajectory}
\newcommand{\rts}{Regge trajectories}
\newcommand{\tr}{trajectory}
\newcommand{\trs}{trajectories}
\newcommand{\bfr}{{\bf r}}
\newcommand{\bfp}{{\bf p}}
\newcommand{\bfpa}{{|\bf p|}}
\newcommand{\gev}{{\rm GeV}}

\newcommand{\cltb}{$\bar{3}_c$}
\newcommand{\cltba}{\bar{3}_c}

\newcommand{\qp}{{q^{\prime}}}

\newcommand{\qpp}{{\bar{q}^{\prime\prime}}}

\newcommand{\qqs}{{[qq^{\prime}]}}
\newcommand{\qqb}{\{qq^{\prime}\}}

\newcommand{\dhts}{doubly heavy triquarks}
\newcommand{\dht}{doubly heavy triquark}

\begin{document}
\title{Regge trajectories for the doubly heavy triquarks $((Qq)\bar{Q}')$}
\author{Xin-Ru Liu}
\email{1170394732@qq.com}
\affiliation{School of Physics and Electronic Engineering, Shanxi Normal University, Taiyuan 030031, China}
\author{Qi Liu}
\email{18803429267@163.com}
\affiliation{School of Physics and Electronic Engineering, Shanxi Normal University, Taiyuan 030031, China}
\author{He Song}
\email{songhe\_22@163.com}
\affiliation{School of Physics and Electronic Engineering, Shanxi Normal University, Taiyuan 030031, China}
\author{Jiao-Kai Chen}
\email{chenjk@sxnu.edu.cn, chenjkphy@outlook.com (corresponding author)}
\affiliation{School of Physics and Electronic Engineering, Shanxi Normal University, Taiyuan 030031, China}

\begin{abstract}
We attempt to apply the Regge trajectory approach to the doubly heavy triquarks $((Qq)\bar{Q}^{\prime})$ $(Q,\,Q'=b,\,c; q=u,\,d,\,s)$. We propose the Regge trajectory relations for the doubly heavy triquarks, and then employ them to crudely estimate the spectra of the triquarks $((cu)\bar{c})$, $((cu)\bar{b})$, $((cs)\bar{c})$, $((cs)\bar{b})$, $((bu)\bar{c})$, $((bu)\bar{b})$, $((bs)\bar{c})$, and $((bs)\bar{b})$. The $\lambda$-trajectories and the $\rho$-trajectories are investigated.
The triquark Regge trajectory becomes a new and very simple approach for estimating the spectra of triquarks. It also provides a simple method to investigate the $\rho$-mode and $\sigma$-mode excitations of pentaquarks and hexaquarks in the triquark picutre. Moreover, the spin-averaged masses of the ground states of pentaquarks $(\bar{c}(cu))(cu)$, $(\bar{b}(bu))(bu)$ and $(\bar{c}(cu))(bu)$ are estimated, which are consistent with other theoretical predictions.
\end{abstract}

\maketitle


\section{Introduction}

Phenomenology suggests that triquark correlations are very important in spectroscopy and for understanding structure of exotic hadrons \cite{Karliner:2003dt,Karliner:2003sy,Lebed:2015tna,Giron:2019bcs,Giron:2021fnl,
Giron:2021sla,Jaffe:2005md,Yasui:2007dv,Hogaasen:2004pm,
Hogaasen:2004ij,Majee:2007gi,JimenezDelgado:2004rd,Lee:2004dp,Zhu:2015bba,Li:2005rb,
Wang:2016dzu,Liu:2019zoy,Wang:2005ms,Jing:2025iqs,Andrew:2023aes,Zhang:2025vqg}.
In the triquark picture, triquarks are constituents of tetraquarks \cite{Yasui:2007dv}, pentaquarks \cite{Karliner:2003dt,Karliner:2003sy,Lebed:2015tna,
Giron:2019bcs,Giron:2021sla,Giron:2021fnl,Hogaasen:2004pm,Hogaasen:2004ij,
Majee:2007gi,JimenezDelgado:2004rd,Lee:2004dp,Zhu:2015bba,Li:2005rb,Wang:2016dzu,
Liu:2019zoy,Wang:2005ms,Jing:2025iqs} and hexaquarks \cite{Andrew:2023aes,Zhang:2025vqg}.

Although triquarks are colored states and not physical, the triquark masses have been studied by various methods, such as Karliner-Lipkin model \cite{Karliner:2003dt,Karliner:2003sy,Majee:2007gi}, sum rule \cite{Lebed:2015tna,Wang:2005ms,Zhang:2025vqg}, operator product expansion \cite{Lee:2004dp}, dynamical diquark model \cite{Giron:2019bcs,Giron:2021sla,Giron:2021fnl}, the color magnetic Hamiltonian \cite{Hogaasen:2004pm}, the quark model \cite{Yasui:2007dv}, vertex function equation \cite{Li:2005rb}, and so on.

Besides these methods, the {\rt}\footnote{A {\rt} of bound states is generally expressed as $M=m_R+\beta_x(x+c_0)^{\nu}$ $(x=l,\,n_r)$ \cite{Chen:2022flh,Chen:2021kfw}, where $M$ is the mass of the bound state, $l$ is the orbital angular momentum, and $n_r$ is the radial quantum number. $m_R$ and $\beta_x$ are parameters. For simplicity, plots in the $(M,\,x)$ plane \cite{Xie:2024lfo}, $(M-m_R,\,x)$ plane \cite{Chen:2023cws}, $(M,\,(x+c_0)^{\nu})$ plane \cite{Song:2025cla}, $(M^2,\,x)$ plane \cite{Chen:2018nnr}, $((M-m_R)^2,\,x)$ plane \cite{Chen:2023djq,Chen:2023web} or $((M-m_R)^{1/{\nu}},\,x)$ plane \cite{Xie:2024dfe}, are all commonly referred as Chew-Frautschi plots. {\rts} can be plotted in these various planes. } is one of the effective approaches widely used in the study of hadron spectra
\cite{Chen:2025fyh,Regge:1959mz,Chew:1962eu,Nambu:1978bd,Gross:2022hyw,Brau:2000st,Brodsky:2006uq,Nielsen:2018uyn,
Brisudova:1999ut,Guo:2008he,Ebert:2009ub,Irving:1977ea,Collins:1971ff,
Inopin:1999nf,Afonin:2014nya,MartinContreras:2020cyg,Sergeenko:1994ck,Veseli:1996gy,
Wilczek:2004im,Selem:2006nd,Sonnenschein:2018fph,MartinContreras:2023oqs,
Roper:2024ovj,Burns:2010qq,Dong:2025coi,Pan:2025gqn,Patel:2025bdu,Lodha:2024yfn,
MartinContreras:2025wnh,Patel:2025rsf,Toniato:2025gts,Guo:2024nrf,Diles:2025xot,
Szanyi:2019kkn,Braga:2025lah,Akbar:2020blk,NaikNS:2025smw,Zhang:2026zoz,Petrov:2026gke}.
In Ref. \cite{Song:2024bkj}, we propose the {\rts} for the triply heavy triquarks, then apply the triquark {\rt} to estimating masses of the triply heavy triquarks. In the present work, we attempt to apply {\rt} approach to the {\dhts} $((Qq)\bar{Q}^{\prime})$ $(Q,\,Q'=b,\,c; q=u,\,d,\,s)$. We roughly estimate the masses of triquarks $((cu)\bar{c})$, $((cu)\bar{b})$, $((cs)\bar{c})$, $((cs)\bar{b})$, $((bu)\bar{c})$, $((bu)\bar{b})$, $((bs)\bar{c})$, and $((bs)\bar{b})$ using the proposed triquark {\rts}.
To our knowledge, there has not yet been theoretical studies addressing the {\rts} for the {\dhts}. The data obtained from other approaches are expected to check our results.

The paper is organized as follows: In Section \ref{sec:rgr}, the {\rt} relations for the {\dhts} are obtained from the spinless Salpeter equation. In Section \ref{sec:rtdiquark}, we investigate the {\rts} for the {\dhts}. The conclusions are presented in Section \ref{sec:conc}.

\section{{\rt} relations}\label{sec:rgr}
In this section, by utilizing the diquark {\rts} \cite{Chen:2023cws}, the $\lambda$-trajectory and $\rho$-trajectory relations for the {\dhts} $((Qq)\bar{Q}^{\prime})$ $(Q,\,Q'=b,\,c; q=u,\,d,\,s)$ are proposed.

\subsection{Preliminary}\label{subsec:prelim}
\begin{figure}[!phtb]
\centering
\includegraphics[width=0.25\textheight]{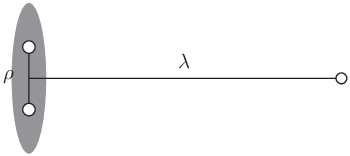}
\caption{Schematic diagram of the doubly heavy triquark $((Qq)\bar{Q}')$ in the antiquark-diquark picture. The grey part represents the heavy-light diquark $(Qq)$ composed of one heavy quark $Q$ and one light quark $q$. The circle on the right denotes the heavy antiquark $\bar{Q}'$.}\label{fig:tr}
\end{figure}

In the diquark picture, a {\dht} $((Qq)\bar{Q}')$ consists of one heavy-light diquark $(Qq)$ and one heavy antiquark $\bar{Q}^{\prime}$. $\lambda$ separates the antiquark $\bar{Q}^{\prime}$ and the diquark $(Qq)$ while $\rho$ separates quark $Q$ and quark $q$ in the diquark $(Qq)$, see Fig. \ref{fig:tr}. There exist two excited modes: the $\rho$-mode involves the radial and orbital excitation in the diquark $(Qq)$, and the $\lambda$-mode involves the radial or orbital excitation between the antiquark $\bar{Q}^{\prime}$ and diquark $(Qq)$. Consequently, there exist two series of {\rts}: one series of $\rho$-{\trs} and one series of $\lambda$-{\trs}.

A quark in $3_c$ and an antiquark in {\cltb} can couple to only two irreducible color representations: $3_c\otimes\cltba=1_c\oplus{8}_c$. The color-Casimir ${\bf{F}_i}\cdot{\bf{F}_j}$ of interaction [see Eq. (\ref{potv})] can be evaluated via the relation \cite{Maiani:2019lpu}
\bea\label{casim}
\langle{\bf{F}_i}\cdot{\bf{F}_j}\rangle=\frac{1}{2}\big[C_2(R)-C_2(R_1)-C_2(R_2)\big],
\eea
where $R_1$ and $R_2$ are the irreducible representations of constituent $1$ and constituent $2$, respectively. $C_2$ is the quadratic Casimir operator. $C_2(R)=C_2(\bar{R})$, $C_2(1_c)=0$, $C_2(3_c)=4/3$, $C_2(6_c)=10/3$, and $C_2(8_c)=3$. For a color singlet meson, $3_c\otimes\cltba\to1_c$. We have the familiar result $\langle{(\bf{F}_i}\cdot{\bf{F}_j})_m\rangle=-4/3$ using Eq. (\ref{casim}), as listed in Table \ref{tab:casimir}.
For a diquark $(Qq)$, $3_c\otimes3_c=\cltba\oplus{6}_c$. We obtain $\langle({\bf{F}_i}\cdot{\bf{F}_j})_d\rangle=-2/3$ for the channel $3_c\otimes3_c\to\cltba$ and $\langle({\bf{F}_i}\cdot{\bf{F}_j})_d\rangle=1/3$ for $3_c\otimes3_c\to{6}_c$. From the signs we see that the internal interaction between the $Qq$ pair is attractive in the $\bar{3}_c$ state and repulsive in the $6_c$ state.
In some works, both $\cltba$ diquark and ${6}_c$ diquark are considered, for example, see Refs. \cite{Maiani:2019lpu,Berwein:2024ztx}, while in other works, only $\cltba$ diquark is considered, for example, see Refs. \cite{Brodsky:2014xia,Galkin:2023wox}. Following Refs. \cite{Brodsky:2014xia,Galkin:2023wox}, only the $\bar{3}_c$ diquark is considered in the present work.

The {\dht} $((Qq)\bar{Q}')$ under consideration is a ${3}_c$ bound state consisting of a constituent diquark $(Qq)$ in the $\cltba$ representation and a constituent antiquark $\bar{Q}'$ also in the $\cltba$ representation. Using Eq. (\ref{casim}), we have $\langle({\bf{F}_i}\cdot{\bf{F}_j})_t\rangle=-2/3$ for $\cltba\otimes\cltba\to3_c$.
In the diquark-antidiquark picture, a color-singlet tetraquark is composed of one diquark in $\cltba$ and one antidiquark in $3_c$. The color Casimir is $\langle({\bf{F}_i}\cdot{\bf{F}_j})_t\rangle=-4/3$ for $\cltba\otimes3_c\to1_c$ \cite{Chen:2023web}.
In the triquark-diquark model, a diquark in $\bar{3}_c$ and a triquark in $3_c$ can form a color-singlet pentaquark \cite{Lebed:2015tna,Song:2025cla}.

\begin{table}[!phtb]
\caption{Values of color-Casimir. Quark $q$, antidiquark $(\bar{q}_1\bar{q}_2)^{{3}_c}$, and triquark $((q_1q_2)\bar{q}_3)^{{3}_c}$ are $3_c$ constituents. Antiquark $\bar{q}$, diquark $({q}_1{q}_2)^{\bar{3}_c}$, and antitriquark $((\bar{q}_1\bar{q}_2){q}_3)^{\bar{3}_c}$ are $\bar{3}_c$ constituents.} \label{tab:casimir}
\centering
\begin{tabular}{ccc}
\hline\hline
          Decomposition        & Channel  &  $\langle{\bf{F}_i}\cdot{\bf{F}_j}\rangle$ \\
\hline\\[-2.5mm]
$3_c\otimes\cltba=1_c\oplus{8}_c$   & $3_c\otimes\cltba\to1_c$   &  $-4/3$  \\
                                    & $3_c\otimes\cltba\to{8}_c$   &  \hspace{2.5mm}${1}/{6}$  \\
$3_c\otimes3_c=\cltba\oplus{6}_c$   & $3_c\otimes3_c\to\cltba$   & $-{2}/{3}$ \\
                                    & $3_c\otimes3_c\to 6_c$   &  \hspace{2.5mm}${1}/{3}$ \\
\hline\hline
\end{tabular}
\end{table}

\begin{table}[!phtb]
\caption{The completely antisymmetric states for the heavy-light diquark $(Qq)$ in {\cltb} \cite{Feng:2023txx}. $j_d$ is the spin of the diquark $(Qq)$, $s_d$ denotes the total spin of quark $Q$ and quark $q$, $l$ represents the orbital angular momentum. $n=n_r+1$, $n_r$ is the radial quantum number, $n_r=0,1,2,\cdots$. $Q$ and $q$ denote the heavy quark and the light quark, respectively.}  \label{tab:dqstates}
\centering
\begin{tabular*}{0.49\textwidth}{@{\extracolsep{\fill}}ccccc@{}}
\hline\hline
 Spin of diquark & Parity  &  Wave state  &  Configuration    \\
( $j_d$ )          & $(j_d^P)$ & $(n^{2s_d+1}l_{j_d})$  \\
\hline
$j_d=0$              & $0^+$   & $n^1s_0$         & $[Qq]^{{\cltba}}_{n^1s_0}$ \\
                 & $0^-$   & $n^3p_0$         & $[Qq]^{{\cltba}}_{n^3p_0}$       \\
$j_d=1$              & $1^+$   & $n^3s_1$, $n^3d_1$   & $\{Qq\}^{{\cltba}}_{n^3s_1}$,\;    $\{Qq\}^{{\cltba}}_{n^3d_1}$\\
                 & $1^-$   & $n^1p_1$, $n^3p_1$   &
$\{Qq\}^{{\cltba}}_{n^1p_1}$,\; $[Qq]^{{\cltba}}_{n^3p_1}$ \\
$j_d=2$              & $2^+$   & $n^1d_2$, $n^3d_2$         &  $[Qq]^{{\cltba}}_{n^1d_2}$,\; $\{Qq\}^{{\cltba}}_{n^3d_2}$\\
                 & $2^-$   & $n^3p_2$, $n^3f_2$       &
 $[Qq]^{{\cltba}}_{n^3p_2}$,\; $[Qq]^{{\cltba}}_{n^3f_2}$       \\
$\cdots$         & $\cdots$ & $\cdots$               & $\cdots$  \\
\hline\hline
\end{tabular*}
\end{table}

In the diquark picture, the state of a {\dht} is denoted as
\bea\label{tetnot}
\left((Qq)^{\cltba}_{n^{2s_d+1}l_{j_d}}\bar{Q}'\right)^{3_c}_{N^{2S+1}L_J}.
\eea
The diquark $(Qq)$ is $\{Qq\}$ or $[Qq]$. $\{Qq\}$ and $[Qq]$ denote the symmetric and antisymmetric flavor wave functions, respectively. The completely antisymmetric states for the diquarks in {\cltb} are listed in Table \ref{tab:dqstates}. $N=N_{r}+1$, where $N_{r}=0,\,1,\,\cdots$. $n=n_{r}+1$, where $n_{r}=0,\,1,\,\cdots$. $N_r$ and $n_{r}$ are the radial quantum numbers of the triquark and diquark, respectively.
$\vec{J}=\vec{L}+\vec{S}$, $\vec{S}=\vec{j}_d+\vec{s}_{\bar{Q}'}$, $\vec{j}_d=\vec{s}_d+\vec{l}$.
$\vec{J}$, $\vec{j}_d$ and $\vec{s}_{\bar{Q}'}$ are the spins of triquark, diquark and antiquark $\bar{Q}'$, respectively. $\vec{S}$ is the summed spin of diquark and antiquark in the triquark. $\vec{s}_{d}$ is the summed spin of quarks in the diquark. $L$ and $l$ are the orbital quantum numbers of triquark and diquark, respectively.

There are two types of {\dhts}: $((Qq)\bar{Q}^{\prime})$ and $((QQ')\bar{q})$, with $Q,Q'\in\{b,\,c\}$ and $q\in\{u,\,d,\,s\}$. $((QQ')\bar{q})$ are discussed in Refs. \cite{Liu:2026npi,Liu:2026btq}. $((Qq)\bar{Q}^{\prime})$ are studied in the present work. For triquark $((Qq)\bar{Q}^{\prime})$, heavy quark $Q$, heavy antiquark $\bar{Q}$, and light quark $q$ are all distinguishable from one another. In contrast, for triquark $((QQ')\bar{q})$, the antisymmetry constraint imposed by identical quarks $Q=Q'$ leads to nonexist states. For instance, $n^3p_2$ for diquark $(QQ)$ do not exist; details can be found in Ref. \cite{Feng:2023txx}.

\subsection{Spinless Salpeter equation}
The spinless Salpeter equation \cite{Godfrey:1985xj,Capstick:1986ter,Ferretti:2019zyh,Bedolla:2019zwg,Durand:1981my,Durand:1983bg,Lichtenberg:1982jp,Jacobs:1986gv} reads as
\begin{eqnarray}\label{qsse}
M\Psi_{d,t}({\bfr})=\left(\omega_1+\omega_2\right)\Psi_{d,t}({\bfr})+V_{d,t}\Psi_{d,t}({\bfr}),
\end{eqnarray}
where $M$ is the bound state mass (diquark or triquark). $\Psi_{d,t}({\bfr})$ are the diquark wave function and the triquark wave function, respectively. $V_{d,t}$ denotes the diquark potential and the triquark potential, respectively (see Eq. (\ref{potv})). $\omega_1$ is the relativistic energy of constituent $1$ (quark or diquark), and $\omega_2$ is of constituent $2$ (quark or antiquark),
\bea\label{omega}
\omega_i=\sqrt{m_i^2+{\bf p}^2}=\sqrt{m_i^2-\Delta}\;\; (i=1,2).
\eea
$m_1$ and $m_2$ are the effective masses of constituent $1$ and $2$, respectively.

Following Refs. \cite{Ferretti:2019zyh,Bedolla:2019zwg,Ferretti:2011zz,Eichten:1974af}, we employ the potential
\begin{align}\label{potv}
V_{d,t}&=-\left[-V_c+\frac{3}{4}{\sigma}r+\frac{3}{4}C\right]
\left({\bf{F}_i}\cdot{\bf{F}_j}\right)_{d,t},
\end{align}
where $\left({\bf{F}_i}\cdot{\bf{F}_j}\right)V_c$ is a color Coulomb potential $(-4/3)(\alpha_s/r)$ or a Coulomb-like potential due to one-gluon-exchange. $\sigma$ is the string tension. $C$ is a fundamental parameter \cite{Gromes:1981cb,Lucha:1991vn}. From Table \ref{tab:casimir}, we have ${\bf{F}_i}\cdot{\bf{F}_j}$ for diquarks and triquarks,
\bea\label{mrcc}
\langle{(\bf{F}_i}\cdot{\bf{F}_j})_{d,t}\rangle=-\frac{2}{3}.
\eea

The effect of the finite size of diquark is treated differently. In Ref. \cite{Faustov:2021hjs}, the size of diquark is taken into account through corresponding form factors. In Ref. \cite{Tourbez:2025hlh}, the diquark size is taken into account by the convoluted potentials. At times, diquark is taken as being pointlike \cite{Ferretti:2019zyh,Lundhammar:2020xvw}.
It is expected that the diquark's finite-size effect does not affect the {\rt} behavior but does make solving process more complex, and sometimes even intractable. In the present work, we focus on triquark {\rts}. For simplicity, the diquark is treated as pointlike.

\subsection{{\rt} relations for heavy-heavy and heavy-light systems}
For the heavy-heavy systems, $m_{1},m_2{\gg}{\bfpa}$, Eq. (\ref{qsse}) reduces to
\begin{eqnarray}\label{qssenrr}
M\Psi_{d,t}({\bfr})=\left[(m_1+m_2)+\frac{{\bfp}^2}{2\mu}+V_{d,t}\right]\Psi_{d,t}({\bfr}),
\end{eqnarray}
where
\bea\label{rdmu}
\mu=m_1m_2/(m_1+m_2).
\eea
By employing the Bohr-Sommerfeld quantization approach \cite{Brau:2000st} and using Eqs. (\ref{potv}) and (\ref{qssenrr}), we obtain the parametrized relation  \cite{Chen:2022flh,Chen:2021kfw}
\bea\label{massform}
M=m_R+\beta_x(x+c_{0x})^{2/3}\,\,(x=l,\,n_r,\,L,\,N_r),
\eea
with
\bea\label{parabm}
\beta_x=c_{fx}c_xc_c,\quad m_R=m_1+m_2+C',
\eea
where
\bea\label{cprime}
C'=C/2,\quad
\sigma'=\sigma/2.
\eea
$c_{x}$ and $c_c$ are
\bea\label{cxcons}
c_c=\left(\frac{\sigma'^2}{\mu}\right)^{1/3},\quad c_{l,L}=\frac{3}{2},\quad c_{n_r,N_r}=\frac{\left(3\pi\right)^{2/3}}{2}.
\eea
$c_{fx}$ are equal theoretically to one and are fitted in practice.
In Eq. (\ref{massform}), $m_1$, $m_2$, $c_x$ and $\sigma$ are universal for the heavy-heavy systems. $c_{0x}$ vary with different {\rts}.

For the heavy-light systems ($m_1\to\infty$ and $m_2\to0$), Eq. (\ref{qsse}) simplifies to
\begin{eqnarray}\label{qssenr}
M\Psi_{d,t}({\bfr})=\left[m_1+{\bfpa}+V_{d,t}\right]\Psi_{d,t}({\bfr}).
\end{eqnarray}
By employing the Bohr-Sommerfeld quantization approach \cite{Brau:2000st} and using Eq. (\ref{qssenr}), the parameterized formula can be written as \cite{Chen:2022flh,Chen:2021kfw}
\bea\label{rtmeson}
M=m_R+\beta_x\sqrt{x+c_{0x}}\;(x=l,\,n_r,\,L,\,N_r).
\eea
$\beta_x$ is in Eq. (\ref{parabm}), and
with
\bea\label{cxcons}
c_{c}=\sqrt{\sigma'},\quad c_{l,L}=2,\quad c_{n_r,N_r}=\sqrt{2\pi}.
\eea
For the heavy-light systems, the common choice of $m_R$ is \cite{Selem:2006nd,Chen:2021kfw,Veseli:1996gy}
\bea\label{mrm1}
m_R=m_1.
\eea

The usual {\rt}, Eq. (\ref{rtmeson}) with (\ref{mrm1}), is obtained in the limit $m_1\to\infty$ and $m_2\to0$.
In Ref. \cite{Chen:2023cws}, we propose two modified formulas to include the light constituent's mass, which can universally describe both the heavy-light mesons and the heavy-light diquarks.
One is Eq. (\ref{rtmeson}) with $m_R$ in (\ref{parabm}), where $m_2$ is the light constituent's mass. Another reads
\bea\label{mrtf}
M=m_R+\sqrt{\beta_x^2(x+c_{0x})+\kappa_{x}m^{3/2}_2(x+c_{0x})^{1/4}}
\eea
if $m_2{\ll}M$, where
\bea\label{mrfp}
m_R=m_1+C',\quad \kappa_x=\frac{4}{3}\sqrt{{\pi}\beta_x},
\eea
where $\beta_x$ is in (\ref{parabm}).
Equation (\ref{rtmeson}) with (\ref{parabm}) is an extension of \cite{Afonin:2014nya}
\bea\label{afoninequ}
M=m_1+m_2+\sqrt{a(n_r+{\alpha}l+b)}
\eea
and the formula \cite{Chen:2022flh}
\bea\label{rmfnpb}
(M-m_1-m_2-C)^2=\alpha_x(x+c_0)^{\gamma}
\eea
while (\ref{mrtf}) with (\ref{mrfp}) is based on the results in \cite{Selem:2006nd,Sonnenschein:2018fph}.
As $m_2=0$, these two modified formulas, Eq. (\ref{rtmeson}) with (\ref{parabm}) and Eq. (\ref{mrtf}) with (\ref{mrfp}), become identical. As $m_2=0$ and $C$ is neglected, these two modified formulas reduce to the usual {\rt} formula for the heavy-light systems, i.e., (\ref{rtmeson}) with (\ref{mrm1}).
Although they give different behavior of $m_2$, Eq. (\ref{rtmeson}) with (\ref{parabm}) and Eq. (\ref{mrtf}) with (\ref{mrfp}) produce consistent results for $l,\,n_r<10$ and have the same behavior $M{\sim}x^{1/2}$ \cite{Chen:2023cws}.

\begin{table}[!phtb]
\caption{The coefficients for the heavy-heavy systems (HHS) and the heavy-light systems (HLS).}  \label{tab:eparam}
\centering
\begin{tabular*}{0.47\textwidth}{@{\extracolsep{\fill}}ccc@{}}
\hline\hline
                   & HHS &  HLS   \\
\hline
$\nu$    & $2/3$ & $1/2$    \\
$c_c$    & $\left({\sigma'^2}/{\mu}\right)^{1/3}$    & $\sqrt{\sigma'}$  \\
$c_{l,L}$    & $3/2$ & $2$   \\
$c_{n_r,N_r}$ & ${\left(3\pi\right)^{2/3}}/{2}$      & $\sqrt{2\pi}$   \\
\hline
\hline
\end{tabular*}
\end{table}

When Eq. (\ref{rtmeson}) with (\ref{parabm}) is employed to discuss the heavy-light systems, summarizing Eqs. (\ref{massform}), (\ref{rtmeson}) with (\ref{parabm}), we have a general form of the {\rts} \cite{Chen:2022flh,Xie:2024lfo}
\begin{align}\label{massfinal}
M=&m_R+\beta_x(x+c_{0x})^{\nu}\,\,(x=l,\,n_r,\,L,\,N_r),\nonumber\\
m_R=&m_1+m_2+C',\quad \beta_x=c_{fx}c_xc_{c},
\end{align}
where ${\nu}$, $c_x$ and $c_{c}$ are listed in Table \ref{tab:eparam}. $c_{fx}$ are theoretically equal to one and are fitted in practice. $c_{0x}$ vary with different {\rts}. Eq. (\ref{massfinal}) can be employed to discuss various systems including the heavy-heavy systems and the heavy-light systems: diquarks, mesons, baryons, triquarks, tetraquarks, and pentaquarks \cite{Chen:2023djq,Chen:2023ngj,Chen:2023web,Chen:2025fyh,Song:2025cla,Song:2024bkj}.

It should be noticed that the general form (\ref{massfinal}) is provisional. Because there are different methods to include the masses of the light constituents. To distinguish the better one, more theoretical and experimental data are needed. In addition to this, the parameter values are universal for both the heavy-heavy systems and the heavy-light systems \cite{Chen:2023cws,Feng:2023txx}, while the parameter values should change for the light systems \cite{Chen:2023ngj} to obtain agreeable results.

\subsection{{\rt} relations for the {\dhts} $((Qq)\bar{Q}')$ }

A {\dht} $((Qq)\bar{Q}')$ consists of one heavy-light diquark and one heavy antiquark, therefore, it is a heavy-heavy system for the $\lambda$-mode. One of the constituents of the {\dht}, the heavy-light diquark composed of one heavy quark and one light quark, is a heavy-light system. Using Eqs. (\ref{massform}), (\ref{rtmeson}), (\ref{parabm}), or (\ref{massfinal}), we have the {\rt} relations for the {\dhts} $((Qq)\bar{Q}')$
\begin{align}\label{t2q}
M&=m_{R_{\lambda}}+\beta_{x_{\lambda}}(x_{\lambda}+c_{0x_{\lambda}})^{2/3}\;(x_{\lambda}=L,\,N_r),\nonumber\\
M_{d}&=m_{R_\rho}+\beta_{x_{\rho}}\sqrt{x_{\rho}+c_{0x_{\rho}}}\;(x_{\rho}=l,\,n_{r}),
\end{align}
where
\begin{align}\label{pa2qQ}
m_{R_{\lambda}}&=M_{d}+m_{Q'}+C/2,\nonumber\\
m_{R_\rho}&=m_{Q}+m_{q}+C/2,\nonumber\\
\beta_{L}&=\frac{3}{2}\left(\frac{\sigma^2}{4\mu_{\lambda}}\right)^{1/3}c_{fL},\; \beta_{N_r}=\frac{(3\pi)^{2/3}}{2}\left(\frac{\sigma^2}{4\mu_{\lambda}}\right)^{1/3}c_{fN_r},\nonumber\\
\mu_{\lambda}&=\frac{M_{d}m_{Q'}}{M_{d}+m_{Q'}},\,
\beta_{l}=\sqrt{2\sigma}c_{fl},\, \beta_{n_r}=\sqrt{\pi\sigma}c_{fn_r}.
\end{align}
In Eq. (\ref{t2q}), $M$ is the triquark mass, and $M_{d}$ is the diquark mass. The second relation in Eq. (\ref{t2q}) is used to calculate the diquark masses \cite{Chen:2023cws}. The relations in Eqs. (\ref{t2q}) and (\ref{pa2qQ}) are employed to calculate the triquark mass.

According to Eqs. (\ref{t2q}) and (\ref{pa2qQ}), we have
\bea\label{lambdform}
M=M_{d}+m_{Q'}+C/2+\beta_{x_{\lambda}}(x_{\lambda}+c_{0x_{\lambda}})^{2/3}
\eea
when the diquark is regarded as a constituent and the structure of the diquark is not considered, then we have the binding energies of the {\dhts}, $\epsilon=C/2+\beta_{x_{\lambda}}(x_{\lambda}+c_{0x_{\lambda}})^{2/3}$. When the diquark is considered as a bound state, we have
\begin{align}\label{combrt}
M=&m_{Q}+m_{q}+m_{Q'}+C\nonumber\\
&+\beta_{x_{\lambda}}(x_{\lambda}+c_{0x_{\lambda}})^{2/3}
+\beta_{x_{\rho}}\sqrt{x_{\rho}+c_{0x_{\rho}}}
\end{align}
from Eqs. (\ref{t2q}) and (\ref{pa2qQ}), then the binding energies of the {\dhts} $((Qq)\bar{Q}')$ read $\epsilon=C+\beta_{x_{\lambda}}(x_{\lambda}+c_{0x_{\lambda}})^{2/3}+
\beta_{x_{\rho}}\sqrt{x_{\rho}+c_{0x_{\rho}}}$.
We can see from Eq. (\ref{combrt}) that there are two series of {\rts} for the {\dhts} $((Qq)\bar{Q}')$: the $\lambda$-trajectories as $x_{\rho}$ is fixed and the $\rho$-trajectories as $x_{\lambda}$ is fixed.

For later convenience, we refer to the {\rts} obtained from Eqs. (\ref{t2q}) and (\ref{pa2qQ}) or from Eqs. (\ref{combrt}) and (\ref{pa2qQ}) as the complete forms of the {\rts}. The obtained constant and the mode under consideration are referred to the main part of the {\rts}.
For the $\rho$-{\trs}, $\beta_{x_{\lambda}}(x_{\lambda}+c_{0x_{\lambda}})^{2/3}$ becomes a function of $x_{\rho}$ (through the dependence in $\beta_{x_{\lambda}}$). Therefore, the main part is
\bea
\widetilde{m}_R+\beta_{x_{\rho}}\sqrt{x_{\rho}+c_{0x_{\rho}}},
\eea
where
\bea
\widetilde{m}_R=m_{Q}+m_{q}+m_{Q'}+C.
\eea
For the $\rho$-{\trs}, the main parts of the {\rts} are not not equal to the complete forms of the {\rts}.
For the $\lambda$-{\trs}, $\beta_{x_{\rho}}\sqrt{x_{\rho}+c_{0x_{\rho}}}$ becomes constant. Therefore, the main part is
\bea
\widetilde{m}_R+\beta_{x_{\lambda}}(x_{\lambda}+c_{0x_{\lambda}})^{2/3},
\eea
where
\bea
\widetilde{m}_R=m_{Q}+m_{q}+m_{Q'}+C+\beta_{x_{\rho}}\sqrt{x_{\rho}+c_{0x_{\rho}}}.
\eea
For the $\lambda$-{\trs}, the main parts are same as the complete forms.

\section{{\rts} for the {\dhts} $((Qq)\bar{Q}')$}\label{sec:rtdiquark}

In this section, the {\rts} for the {\dhts} $((cu)\bar{c})$, $((cu)\bar{b})$, $((cs)\bar{c})$, $((cs)\bar{b})$, $((bu)\bar{c})$, $((bu)\bar{b})$, $((bs)\bar{c})$, and $((bs)\bar{b})$ are investigated by using Eqs. (\ref{t2q}) and (\ref{pa2qQ}) or Eqs. (\ref{combrt}) and (\ref{pa2qQ}).

\subsection{Parameters}

\begin{table}[!phtb]
\caption{The values of parameters \cite{Chen:2023cws,Faustov:2021hjs}.}  \label{tab:parmv}
\centering
\begin{tabular*}{0.47\textwidth}{@{\extracolsep{\fill}}cl@{}}
\hline\hline
          & $m_{u,d}=0.33\; {\gev}$, \; $m_s=0.50\; {\gev}$, \; $m_b=4.88\; {\gev}$  \\
          & $m_c=1.55\; {\gev}$, \; $\sigma=0.18\; {\gev^2}$,\; $C=-0.3\; {\gev}$ \\
$(bu)$    & $c_{fn_{r}}=0.988$,\; $c_{fl}=0.965$  \\
          & $c_{0n_{r}}(1^1s_0)=0.125$,\;$c_{0n_{l}}(1^1s_0)=0.18$\\
          & $c_{0n_{r}}(1^3s_1)=0.155$,\;  $c_{0n_{l}}(1^3s_1)=0.22$\\
$(bs)$    & $c_{fn_{r}}=0.953$,\; $c_{fl}=0.919$  \\
          & $c_{0n_{r}}(1^1s_0)=0.08$,\; $c_{0n_{l}}(1^1s_0)=0.115$\\
          & $c_{0n_{r}}(1^3s_1)=0.11$,\; $c_{0n_{l}}(1^3s_1)=0.16$\\
$(cu)$    & $c_{fn_{r}}=1.000$,\; $c_{fl}=1.038$  \\
          & $c_{0n_{r}}(1^1s_0)=0.065$,\; $c_{0n_{l}}(1^1s_0)=0.095$\\
          & $c_{0n_{r}}(1^3s_1)=0.17$,\; $c_{0n_{l}}(1^3s_1)=0.19$\\
$(cs)$    & $c_{fn_{r}}=1.016$,\; $c_{fl}=1.015$  \\
          & $c_{0n_{r}}(1^1s_0)=0.03$,\; $c_{0n_{l}}(1^1s_0)=0.055$\\
          & $c_{0n_{r}}(1^3s_1)=0.095$, \; $c_{0n_{l}}(1^3s_1)=0.135$\\
\hline
\hline
\end{tabular*}
\end{table}

The parameter values are listed in Table \ref{tab:parmv}. The values of $m_u$, $m_d$, $m_s$, $m_b$, $m_c$, $\sigma$ and $C$ are taken directly from \cite{Faustov:2021hjs}. $c_{fx}$ and $c_{0x}$ for the $\rho$-mode are obtained by fitting the {\rts} for the heavy-light mesons, and then are used to fit the {\rts} for the heavy-light diquarks \cite{Chen:2023cws}.
The parameters for the $\lambda$-mode are determined by the relations \cite{Xie:2024dfe,Liu:2026npi}
\begin{eqnarray}
c_{fL}=&1.116 + 0.013\frac{\mu_{\lambda}}{m_0},\; c_{0L}=0.540- 0.141\frac{\mu_{\lambda}}{m_0}, \nonumber\\
c_{fN_r}=&1.008 + 0.008\frac{\mu_{\lambda}}{m_0}, \;  c_{0N_r}=0.334 - 0.087\frac{\mu_{\lambda}}{m_0},\label{fitcfxnr}
\end{eqnarray}
where $m_0=1$ {\gev}. $\mu_{\lambda}$ is the reduced masses, see Eq. (\ref{pa2qQ}).

\subsection{$\rho$-{\trs} for the {\dhts} } \label{subsec:rho}

\begin{table*}[!htbp]
\caption{The spin-averaged masses of the $\rho$-excited states of the {\dhts} (in ${\gev}$). The notation in Eq. (\ref{tetnot}) is rewritten as $|n^{2s_d+1}l_{j_d},N^{2S+1}L_J\rangle$. And $|n^{2s_d+1}l_{j_d},NL\rangle$ denotes the spin-averaged states. Eqs. (\ref{t2q}), (\ref{pa2qQ}) and (\ref{fitcfxnr}) [or Eqs. (\ref{combrt}), (\ref{pa2qQ}), and (\ref{fitcfxnr})] are used.}  \label{tab:massrho}
\centering
\begin{tabular*}{1.0\textwidth}{@{\extracolsep{\fill}}ccccccccc@{}}
\hline\hline
\addlinespace[2pt]
  $|n^{2s_d+1}l_{j_d},NL\rangle$ & $((cu)\bar{c})$  & $((cs)\bar{c})$
  & $((cu)\bar{b})$  & $((cs)\bar{b})$ & $((bu)\bar{c})$  & $((bs)\bar{c})$
  & $((bu)\bar{b})$  & $((bs)\bar{b})$ \\
\hline
 $|1^3s_1, 1S\rangle$  &3.63  &3.73  &6.91  &7.00  &6.91  &7.03  &10.16  &10.27     \\
 $|2^3s_1, 1S\rangle$  &4.13  &4.28  &7.40  &7.55  &7.42  &7.54  &10.66  &10.79     \\
 $|3^3s_1, 1S\rangle$  &4.42  &4.58  &7.69  &7.85  &7.71  &7.83  &10.95  &11.07     \\
 $|4^3s_1, 1S\rangle$  &4.65  &4.82  &7.91  &8.08  &7.94  &8.05  &11.18  &11.29     \\
 $|5^3s_1, 1S\rangle$  &4.84  &5.02  &8.11  &8.28  &8.13  &8.24  &11.37  &11.48     \\
 $|1^1s_0, 1S\rangle$  &3.52  &3.62  &6.80  &6.91  &6.88  &6.99  &10.13  &10.24     \\
 $|2^1s_0, 1S\rangle$  &4.09  &4.26  &7.36  &7.53  &7.41  &7.53  &10.65  &10.78     \\
 $|3^1s_0, 1S\rangle$  &4.39  &4.57  &7.66  &7.83  &7.70  &7.82  &10.94  &11.06     \\
 $|4^1s_0, 1S\rangle$  &4.62  &4.80  &7.89  &8.07  &7.93  &8.04  &11.17  &11.28     \\
 $|5^1s_0, 1S\rangle$  &4.82  &5.01  &8.09  &8.27  &8.13  &8.23  &11.36  &11.47     \\
\hline
 $|1^3s_1, 1S\rangle$  &3.59  &3.71  &6.88  &6.99  &6.89  &7.01  &10.14  &10.26     \\
 $|1^3p_2, 1S\rangle$  &3.99  &4.13  &7.27  &7.41  &7.26  &7.38  &10.50  &10.63     \\
 $|1^3d_3, 1S\rangle$  &4.23  &4.37  &7.51  &7.64  &7.48  &7.60  &10.72  &10.84     \\
 $|1^3f_4, 1S\rangle$  &4.42  &4.56  &7.69  &7.82  &7.66  &7.77  &10.90  &11.01     \\
 $|1^3g_5, 1S\rangle$  &4.58  &4.71  &7.85  &7.98  &7.81  &7.91  &11.05  &11.15     \\
 $|1^3h_6, 1S\rangle$  &4.72  &4.85  &7.99  &8.12  &7.94  &8.04  &11.18  &11.28     \\
 $|1^1s_0, 1S\rangle$  &3.52  &3.63  &6.80  &6.92  &6.87  &6.98  &10.11  &10.22     \\
 $|1^1p_1, 1S\rangle$  &3.97  &4.11  &7.24  &7.38  &7.25  &7.37  &10.49  &10.62     \\
 $|1^1d_2, 1S\rangle$  &4.21  &4.35  &7.49  &7.62  &7.47  &7.59  &10.72  &10.83     \\
 $|1^1f_3, 1S\rangle$  &4.40  &4.54  &7.68  &7.81  &7.65  &7.76  &10.89  &11.00     \\
 $|1^1g_4, 1S\rangle$  &4.57  &4.70  &7.84  &7.97  &7.80  &7.91  &11.04  &11.15     \\
 $|1^1h_5, 1S\rangle$  &4.71  &4.84  &7.98  &8.11  &7.93  &8.03  &11.18  &11.27     \\
\hline\hline
\end{tabular*}
\end{table*}

\begin{figure*}[!phtb]
\centering
\subfigure[]{\label{subfigure:cfa}\includegraphics[scale=0.3]{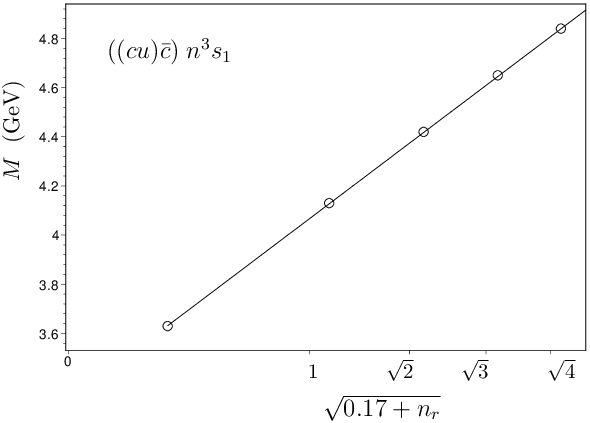}}
\subfigure[]{\label{subfigure:cfa}\includegraphics[scale=0.3]{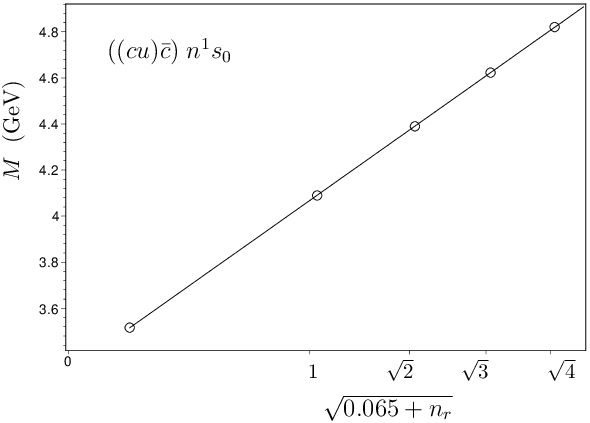}}
\subfigure[]{\label{subfigure:cfa}\includegraphics[scale=0.3]{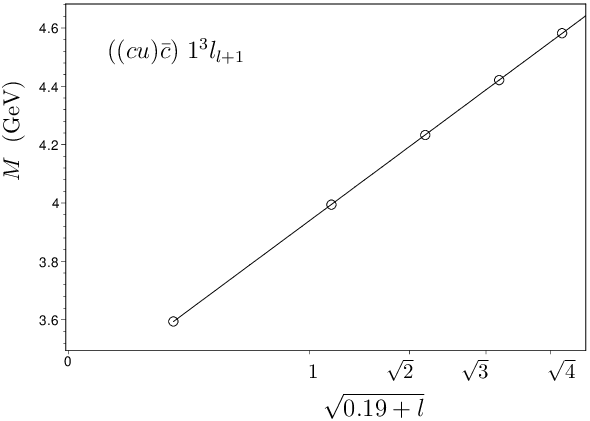}}
\subfigure[]{\label{subfigure:cfa}\includegraphics[scale=0.3]{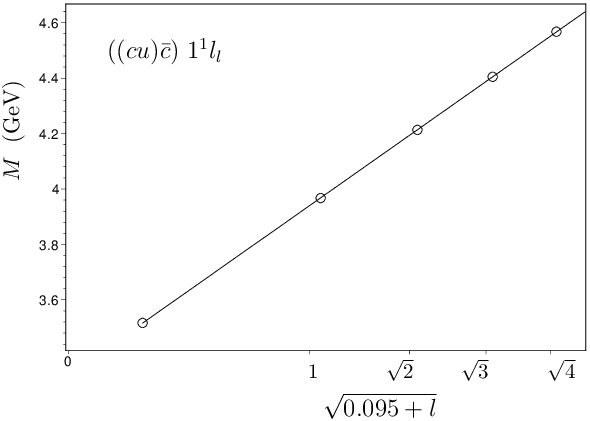}}
\subfigure[]{\label{subfigure:cfa}\includegraphics[scale=0.3]{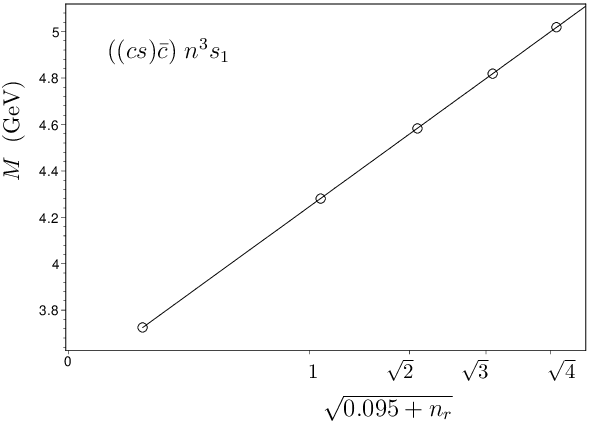}}
\subfigure[]{\label{subfigure:cfa}\includegraphics[scale=0.3]{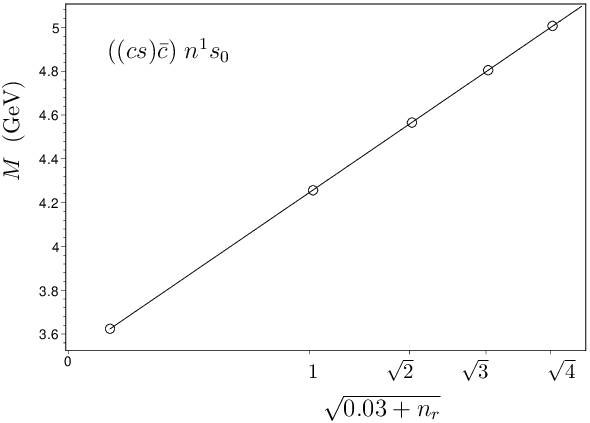}}
\subfigure[]{\label{subfigure:cfa}\includegraphics[scale=0.3]{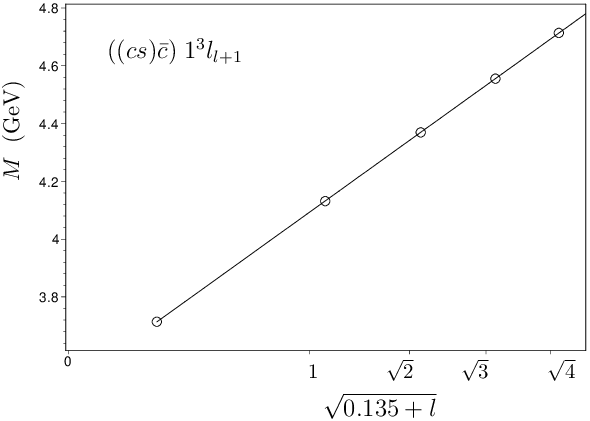}}
\subfigure[]{\label{subfigure:cfa}\includegraphics[scale=0.3]{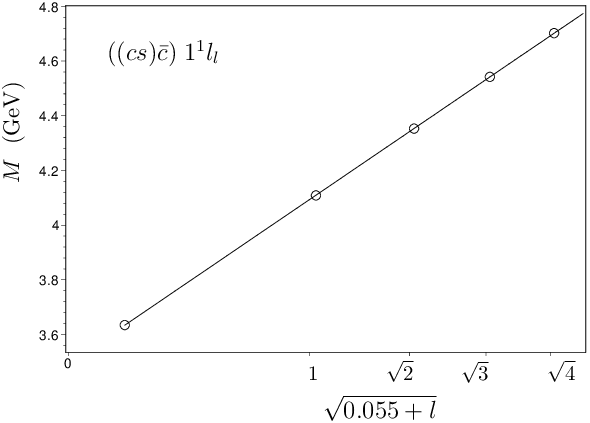}}
\subfigure[]{\label{subfigure:cfa}\includegraphics[scale=0.3]{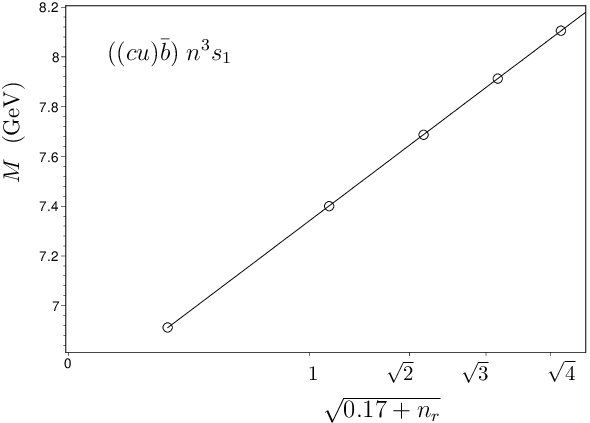}}
\subfigure[]{\label{subfigure:cfa}\includegraphics[scale=0.3]{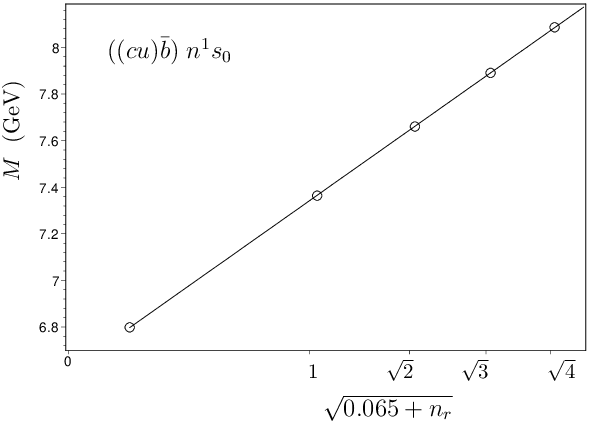}}
\subfigure[]{\label{subfigure:cfa}\includegraphics[scale=0.3]{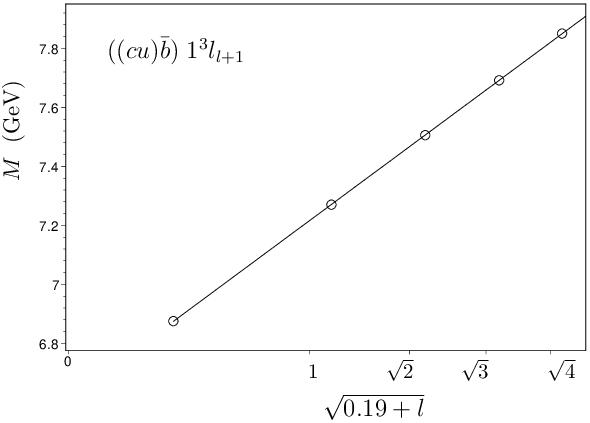}}
\subfigure[]{\label{subfigure:cfa}\includegraphics[scale=0.3]{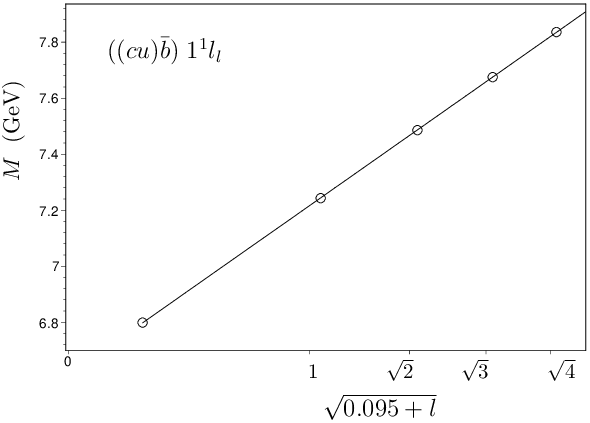}}
\subfigure[]{\label{subfigure:cfa}\includegraphics[scale=0.3]{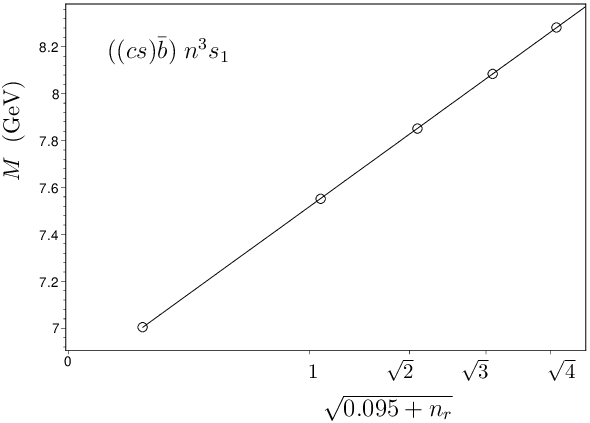}}
\subfigure[]{\label{subfigure:cfa}\includegraphics[scale=0.3]{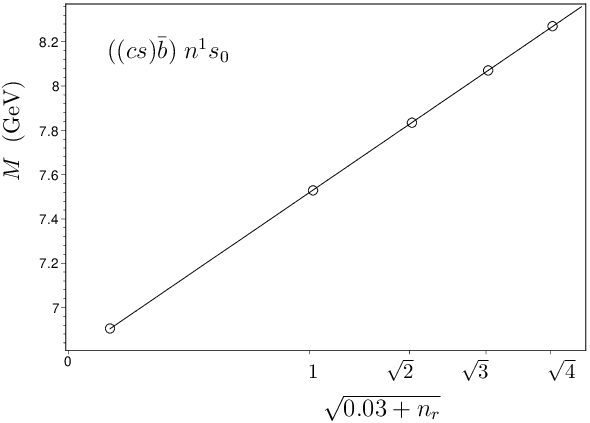}}
\subfigure[]{\label{subfigure:cfa}\includegraphics[scale=0.3]{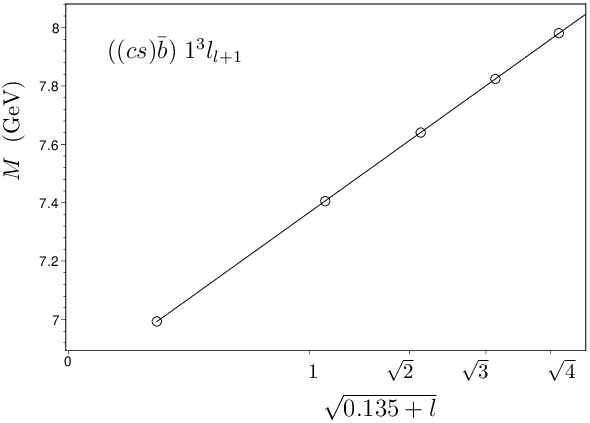}}
\subfigure[]{\label{subfigure:cfa}\includegraphics[scale=0.3]{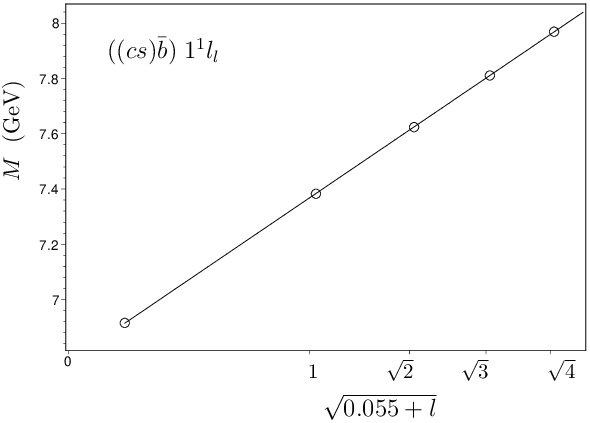}}
\subfigure[]{\label{subfigure:cfa}\includegraphics[scale=0.3]{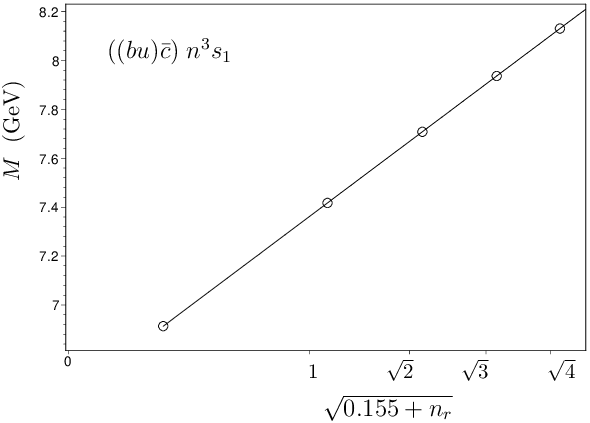}}
\subfigure[]{\label{subfigure:cfa}\includegraphics[scale=0.3]{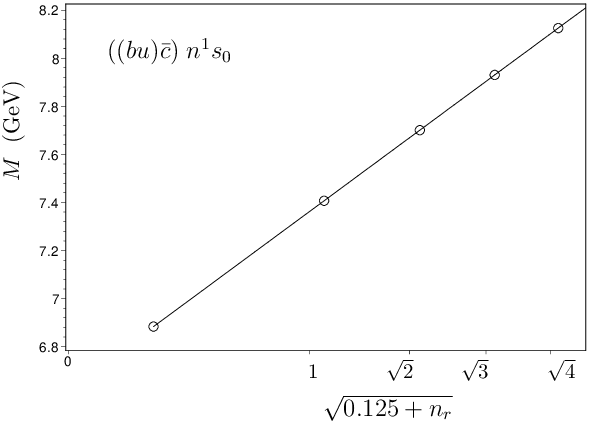}}
\subfigure[]{\label{subfigure:cfa}\includegraphics[scale=0.3]{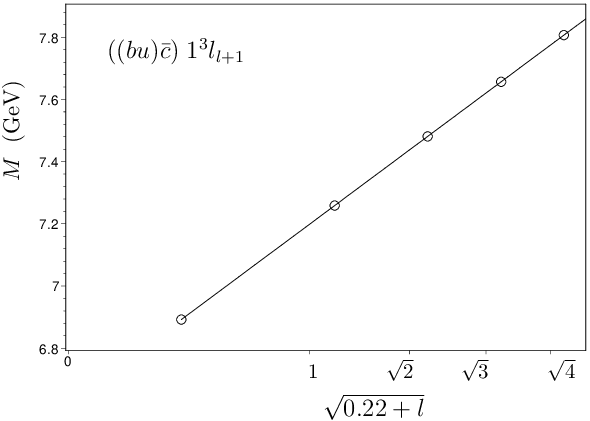}}
\subfigure[]{\label{subfigure:cfa}\includegraphics[scale=0.3]{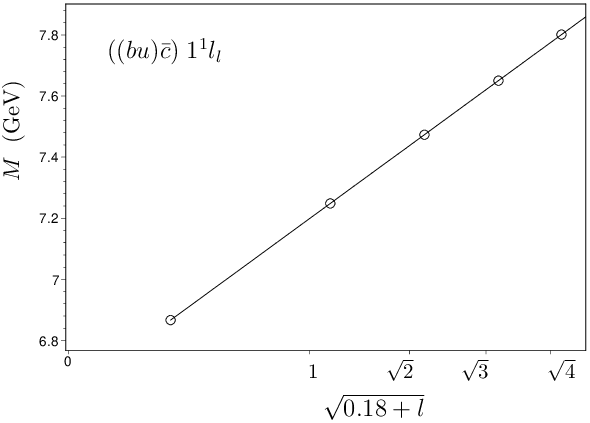}}
\subfigure[]{\label{subfigure:cfa}\includegraphics[scale=0.3]{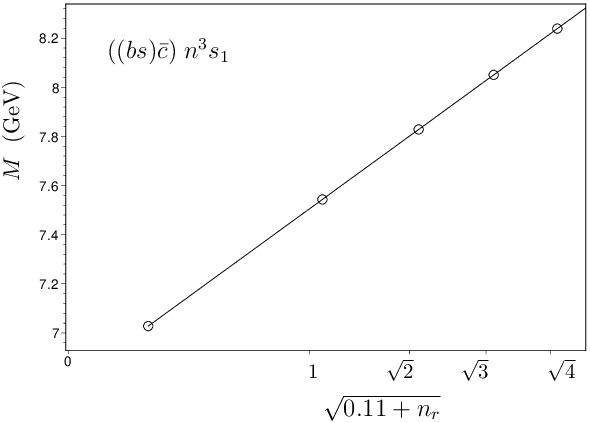}}
\subfigure[]{\label{subfigure:cfa}\includegraphics[scale=0.3]{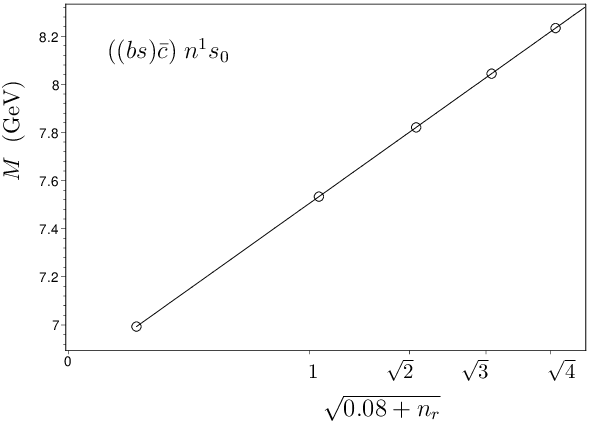}}
\subfigure[]{\label{subfigure:cfa}\includegraphics[scale=0.3]{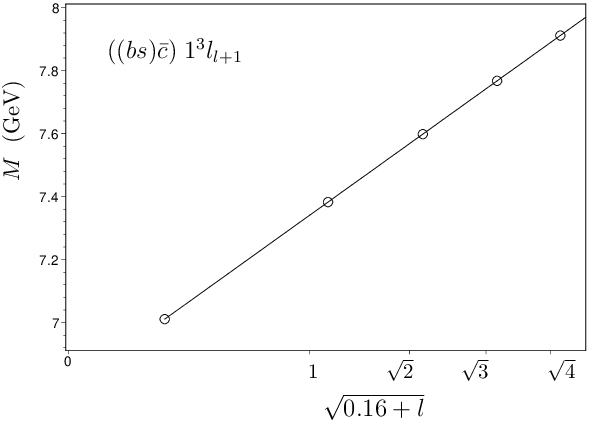}}
\subfigure[]{\label{subfigure:cfa}\includegraphics[scale=0.3]{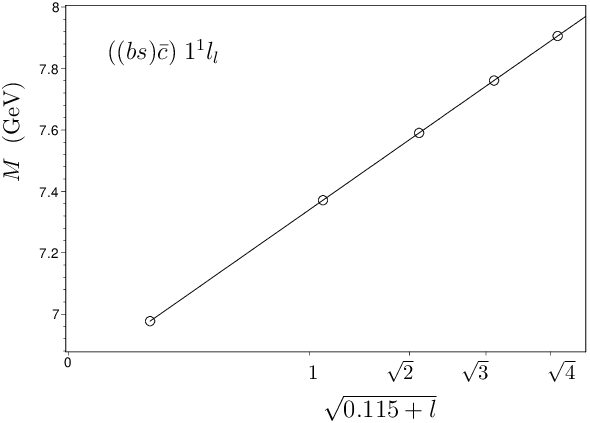}}
\subfigure[]{\label{subfigure:cfa}\includegraphics[scale=0.3]{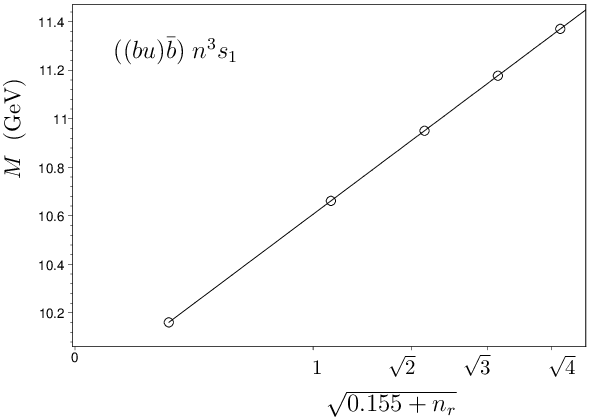}}
\subfigure[]{\label{subfigure:cfa}\includegraphics[scale=0.3]{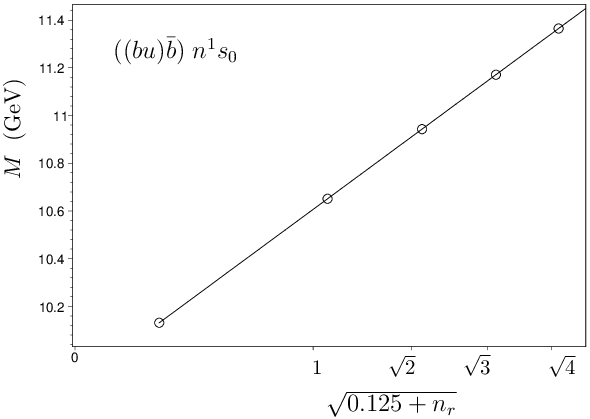}}
\subfigure[]{\label{subfigure:cfa}\includegraphics[scale=0.3]{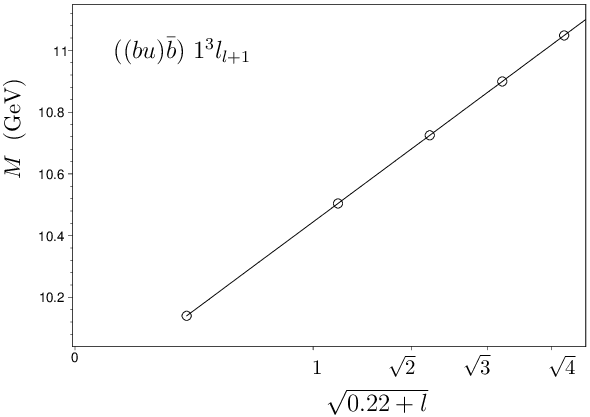}}
\subfigure[]{\label{subfigure:cfa}\includegraphics[scale=0.3]{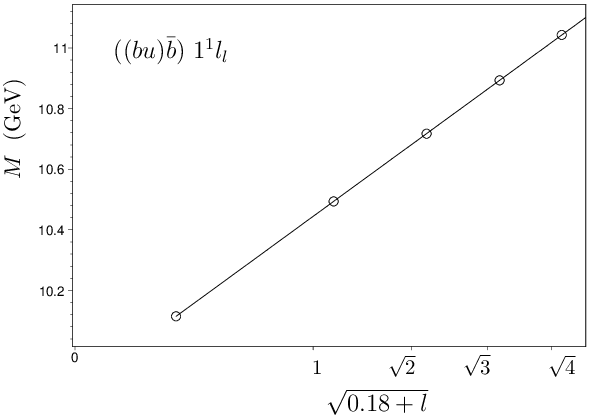}}
\subfigure[]{\label{subfigure:cfa}\includegraphics[scale=0.3]{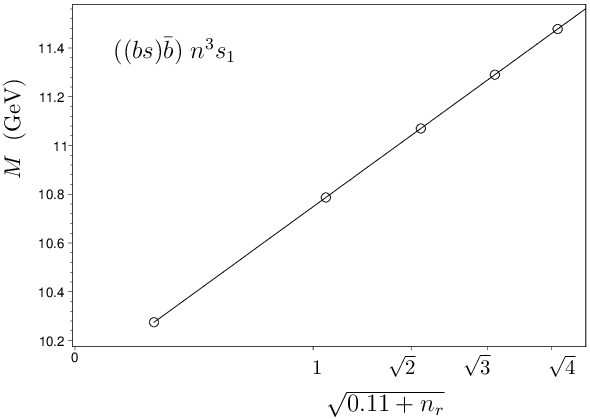}}
\subfigure[]{\label{subfigure:cfa}\includegraphics[scale=0.3]{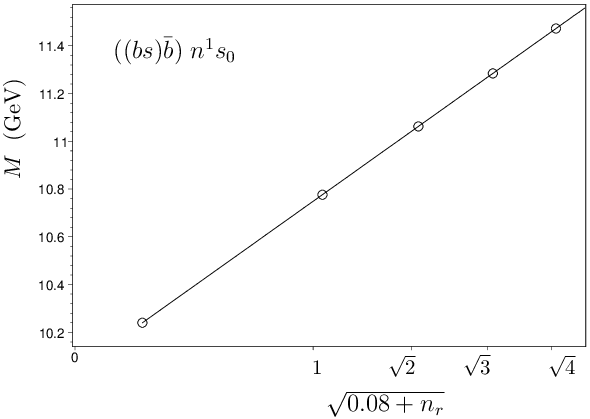}}
\subfigure[]{\label{subfigure:cfa}\includegraphics[scale=0.3]{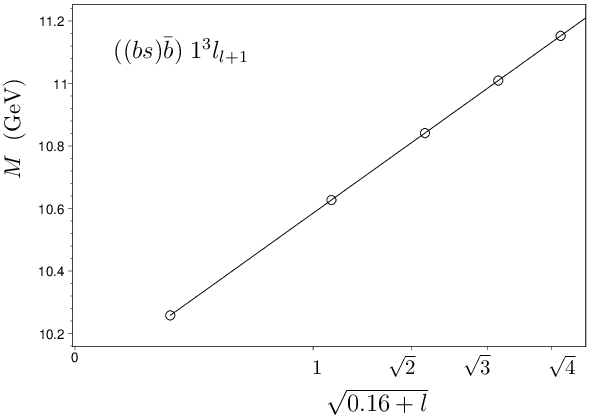}}
\subfigure[]{\label{subfigure:cfa}\includegraphics[scale=0.3]{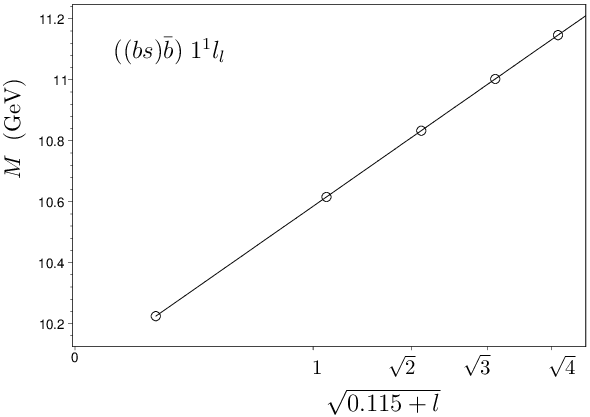}}
\caption{Radial and orbital $\rho$-{\trs} for the {\dhts} $((cu)\bar{c})$, $((cu)\bar{b})$, $((cs)\bar{c})$, $((cs)\bar{b})$, $((bu)\bar{c})$, $((bu)\bar{b})$, $((bs)\bar{c})$, and $((bs)\bar{b})$. $n^1s_0$ and $1^1l_{l}$ denote the spin singlets of diquarks. $n^3s_1$ and $1^3l_{l+1}$ denote the spin triplets. Circles represent the predicted data and the black lines represent the fitted formulas. Data are listed in Table \ref{tab:massrho} and formulas are in Table \ref{tab:rtbehav}.}\label{fig:bc}
\end{figure*}

Employing Eqs. (\ref{t2q}), (\ref{pa2qQ}), (\ref{fitcfxnr}) [or Eqs. (\ref{combrt}), (\ref{pa2qQ}), (\ref{fitcfxnr})], and parameters in Table \ref{tab:parmv}, the spectra of the $\rho$-excited states of the {\dhts} are calculated, see Table \ref{tab:massrho}. The $\rho$-trajectories for the {\dhts} are shown in Fig. \ref{fig:bc}. When calculating the $\rho$-mode radially and orbitally excited states, the $\lambda$-mode state is taken as the radial ground state.

Observing Tables \ref{tab:masslambdac} and \ref{tab:massrho}, we find that the masses of the $\rho$-excited states are greater than that of the $\lambda$-excited states. That is because the binding energy of a heavy-light diquark is greater than the binding energy between a heavy diquark and a heavy antiquark.
Accordingly, the following inequalities hold for the {\dhts}:
\begin{align}\label{masseq}
M\left(\rho, ((Qq)\bar{Q}')\right)>
M\left(\lambda, ((Qq)\bar{Q}')\right),
\end{align}
where $Q,\,Q'=b,c$, $q=u,d,s$. $M\left(\rho, ((Qq)\bar{Q}')\right)$ and $M\left(\lambda, ((Qq)\bar{Q}')\right)$ denote the masses of the radially or orbitally $\rho$-mode and $\lambda$-mode excited state of $((Qq)\bar{Q}')$, respectively.

Unlike the {\rts} for light mesons, which have a simple linear form, the $\rho$-{\trs} for {\dhts} become tedious when plotted in the $(M^2,x)$ plane. For example, for the states $|n^3s_1, 1S\rangle$ of triquark $((cu)\bar{c})$, the fitted $\rho$-{\tr} changes from Eq. (\ref{fra}) to
\bea\label{msqr}
M^2=11.1536+ 0.549176 n_r + 4.92911 \sqrt{0.17+ n_r}.
\eea
The squared complete form of $\rho$-{\tr} [see Eq. (\ref{fulla})] will be even more complicated.
Furthermore, the behavior of the Regge trajectory in terms of $M^2$ is less obvious and transparent than in terms of $M$. For the fitted $\rho$-{\tr}, $M{\sim}\sqrt{n_{r}}$, while $M^2{\sim}n_{r}^{\nu}$ with $\nu$ not being clearly defined but greater than $1/2$ and less than $1$, and $\nu$ varies for different {\trs}. When Eq. (\ref{msqr}) is approximated as $M^2=m'_R+\beta'\sqrt{x+c_0}$ because the term $\sqrt{0.17+ n_r}$ plays the dominant role, the resulting estimates  become less accurate compared with those obtained using $M=m_R+\beta\sqrt{x+c_0}$.

Inspired by Ref. \cite{Burns:2010qq}, we employ the $(M,\,\sqrt{x+c_0})$ plane, rather than the $(M^2,\,x)$ plane, to plot the $\rho$-{\trs} for the triquarks (see Fig. \ref{fig:bc}). The full forms of the $\rho$-{\trs}, directly calculated from Eqs. (\ref{t2q}), (\ref{pa2qQ}), and (\ref{fitcfxnr}), or Eqs. (\ref{combrt}), (\ref{pa2qQ}), and (\ref{fitcfxnr}), are indicated by the circles in Fig. \ref{fig:bc}. For illustration, the full expression for the states $|n^3s_1, 1S\rangle$ of triquark $((cu)\bar{c})$ is presented in Eq. (\ref{fulla}). The full expressions are generally lengthy and cumbersome, and their complexity varies with different {\rts}.
By performing a linear fit in the $(M,\sqrt{c0+ x})$ plane to the calculated data in Table \ref{tab:massrho}, we obtain the fitted formulas listed in Table \ref{tab:rtbehav}. In Fig. \ref{fig:bc}, the fitted formulas are represented by the black lines.
From Fig. \ref{fig:bc}, we can see that the data calculated by the complex full forms of the $\rho$-{\trs} can be well approximated by the simple fitted formulas. Accordingly, the complete forms of the $\rho$-{\trs} exhibit the behavior $M{\sim}\sqrt{x_{\rho}}$, where $x_{\rho}=n_{r},\;l$.
The main parts of the full forms of the $\rho$-{\trs} have the same behavior as the complete forms, but show significant deviations from the full forms, see Eqs. (\ref{fulla})--(\ref{fra}).
In conclusion, the $\rho$-{\trs} are presented in Fig. \ref{fig:bc}, and the Chew-Frautschi plots clearly demonstrate the behavior of {\trs} (see the footnote on the first page). The tedious complete forms of the $\rho$-{\trs}--in which the {\rt} behavior is not obvious--can be well approximated by the simple fitted forms, in which the {\rt} behavior is apparent.

\subsection{$\lambda$-{\trs} for the {\dhts} }\label{subsec:lamb}

\begin{table*}[!phtb]
\caption{Same as Table \ref{tab:massrho} except for the $\lambda$-excited states.}  \label{tab:masslambdac}
\centering
\begin{tabular*}{1.0\textwidth}{@{\extracolsep{\fill}}ccccccccc@{}}
\hline\hline
\addlinespace[2pt]
  $|n^{2s_d+1}l_{j_d},NL\rangle$ & $((cu)\bar{c})$  & $((cs)\bar{c})$
  & $((cu)\bar{b})$  & $((cs)\bar{b})$ & $((bu)\bar{c})$  & $((bs)\bar{c})$
  & $((bu)\bar{b})$  & $((bs)\bar{b})$ \\
\hline
 $|1^3s_1, 1S\rangle$  &3.63  &3.73  &6.91  &7.00  &6.91  &7.03  &10.16  &10.27     \\
 $|1^3s_1, 2S\rangle$  &3.99  &4.08  &7.23  &7.32  &7.24  &7.36  &10.44  &10.56     \\
 $|1^3s_1, 3S\rangle$  &4.26  &4.35  &7.46  &7.54  &7.48  &7.60  &10.64  &10.75     \\
 $|1^3s_1, 4S\rangle$  &4.48  &4.57  &7.65  &7.74  &7.69  &7.80  &10.80  &10.91     \\
 $|1^3s_1, 5S\rangle$  &4.69  &4.77  &7.82  &7.91  &7.87  &7.99  &10.95  &11.06     \\
 $|1^1s_0, 1S\rangle$  &3.52  &3.62  &6.80  &6.91  &6.88  &6.99  &10.13  &10.24     \\
 $|1^1s_0, 2S\rangle$  &3.88  &3.99  &7.12  &7.22  &7.22  &7.32  &10.42  &10.52     \\
 $|1^1s_0, 3S\rangle$  &4.15  &4.25  &7.35  &7.45  &7.45  &7.56  &10.61  &10.72     \\
 $|1^1s_0, 4S\rangle$  &4.37  &4.48  &7.55  &7.64  &7.66  &7.77  &10.77  &10.88     \\
 $|1^1s_0, 5S\rangle$  &4.58  &4.68  &7.72  &7.82  &7.84  &7.95  &10.92  &11.03     \\
\hline
 $|1^3s_1, 1S\rangle$  &3.64  &3.73  &6.92  &7.01  &6.92  &7.03  &10.16  &10.28     \\
 $|1^3s_1, 1P\rangle$  &3.89  &3.98  &7.14  &7.23  &7.15  &7.26  &10.37  &10.48     \\
 $|1^3s_1, 1D\rangle$  &4.08  &4.17  &7.30  &7.39  &7.32  &7.44  &10.51  &10.62     \\
 $|1^3s_1, 1F\rangle$  &4.24  &4.33  &7.45  &7.53  &7.47  &7.59  &10.63  &10.74     \\
 $|1^3s_1, 1G\rangle$  &4.39  &4.48  &7.58  &7.66  &7.61  &7.72  &10.74  &10.85     \\
 $|1^3s_1, 1H\rangle$  &4.53  &4.62  &7.69  &7.78  &7.74  &7.85  &10.84  &10.95     \\
 $|1^1s_0, 1S\rangle$  &3.52  &3.63  &6.80  &6.91  &6.89  &7.00  &10.13  &10.24     \\
 $|1^1s_0, 1P\rangle$  &3.77  &3.88  &7.03  &7.13  &7.12  &7.23  &10.34  &10.44     \\
 $|1^1s_0, 1D\rangle$  &3.97  &4.07  &7.19  &7.30  &7.29  &7.40  &10.48  &10.59     \\
 $|1^1s_0, 1F\rangle$  &4.13  &4.24  &7.34  &7.44  &7.44  &7.55  &10.60  &10.71     \\
 $|1^1s_0, 1G\rangle$  &4.28  &4.39  &7.47  &7.57  &7.58  &7.69  &10.71  &10.82     \\
 $|1^1s_0, 1H\rangle$  &4.42  &4.53  &7.59  &7.69  &7.71  &7.81  &10.81  &10.92     \\
\hline\hline
\end{tabular*}
\end{table*}

\begin{figure*}[!phtb]
\centering
\subfigure[]{\label{subfigure:cfa}\includegraphics[scale=0.3]{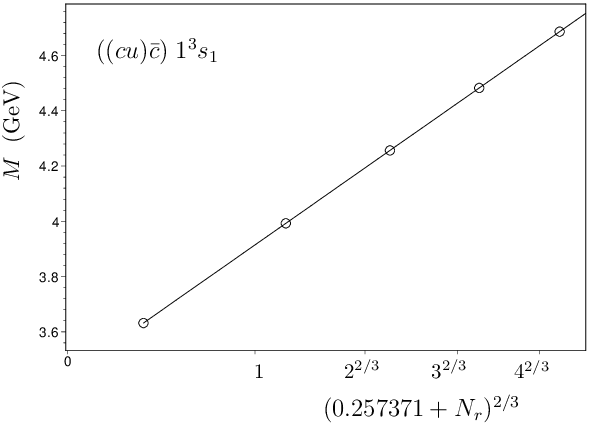}}
\subfigure[]{\label{subfigure:cfa}\includegraphics[scale=0.3]{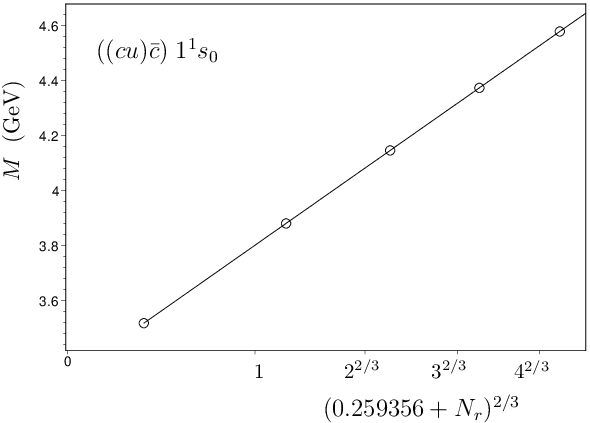}}
\subfigure[]{\label{subfigure:cfa}\includegraphics[scale=0.3]{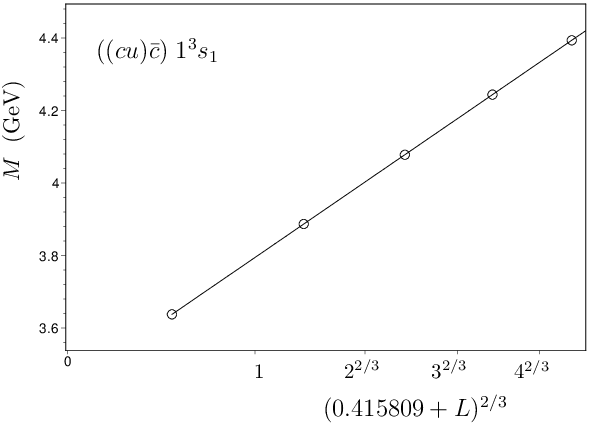}}
\subfigure[]{\label{subfigure:cfa}\includegraphics[scale=0.3]{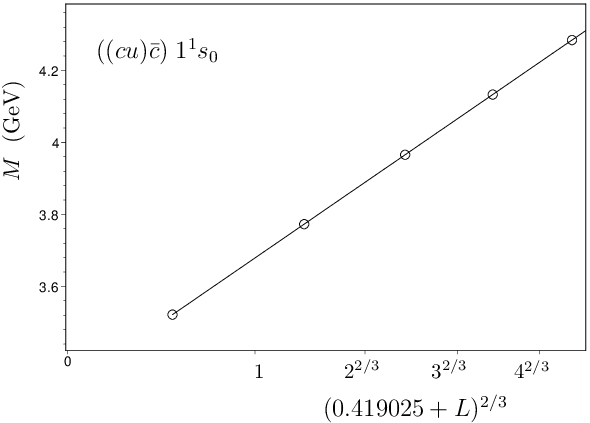}}
\subfigure[]{\label{subfigure:cfa}\includegraphics[scale=0.3]{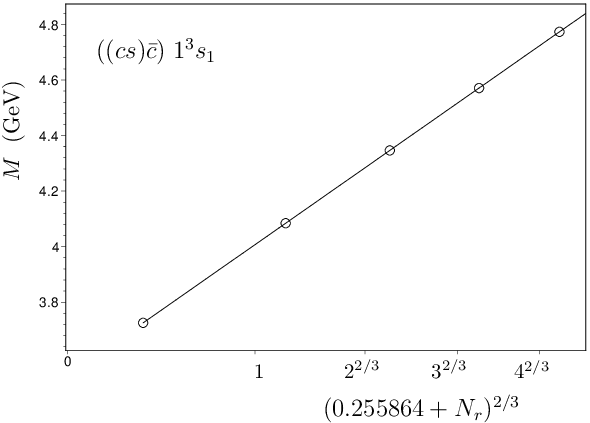}}
\subfigure[]{\label{subfigure:cfa}\includegraphics[scale=0.3]{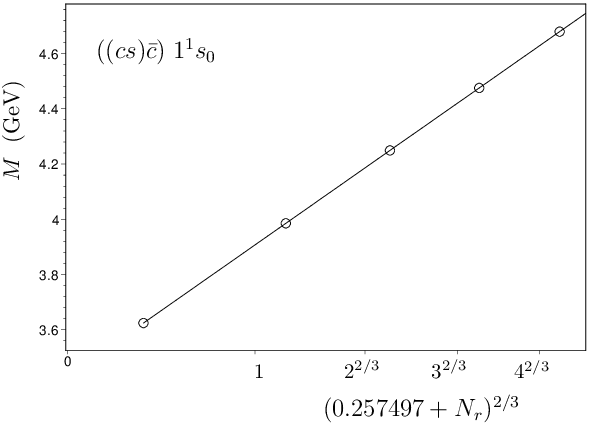}}
\subfigure[]{\label{subfigure:cfa}\includegraphics[scale=0.3]{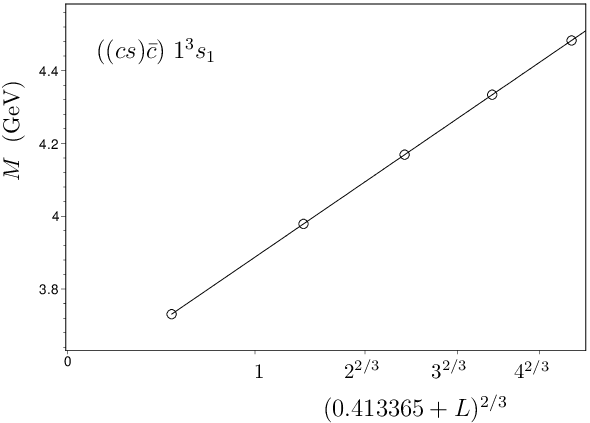}}
\subfigure[]{\label{subfigure:cfa}\includegraphics[scale=0.3]{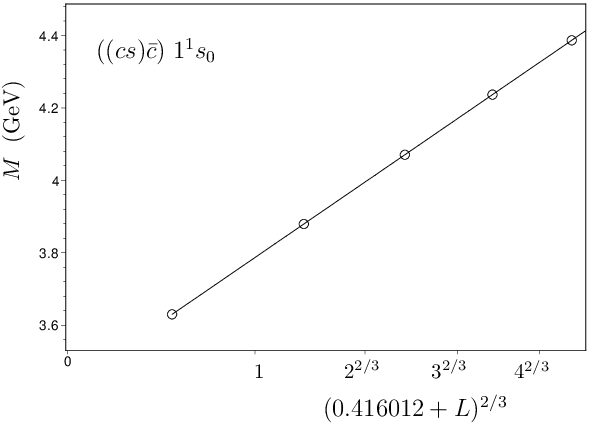}}
\subfigure[]{\label{subfigure:cfa}\includegraphics[scale=0.3]{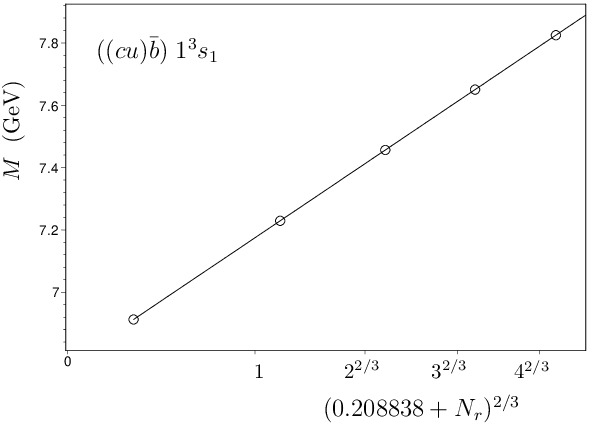}}
\subfigure[]{\label{subfigure:cfa}\includegraphics[scale=0.3]{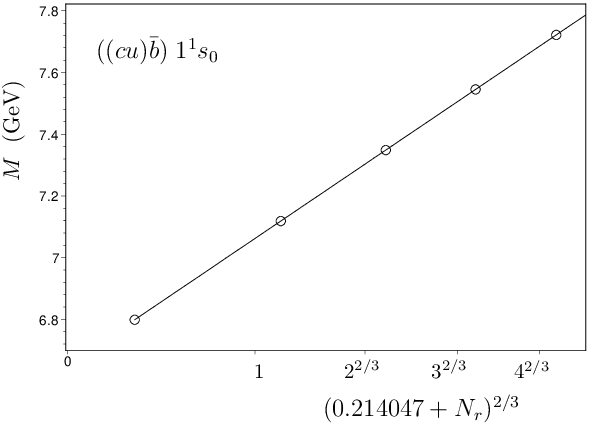}}
\subfigure[]{\label{subfigure:cfa}\includegraphics[scale=0.3]{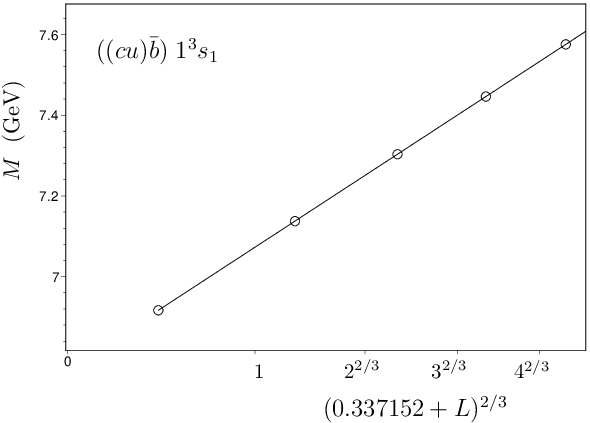}}
\subfigure[]{\label{subfigure:cfa}\includegraphics[scale=0.3]{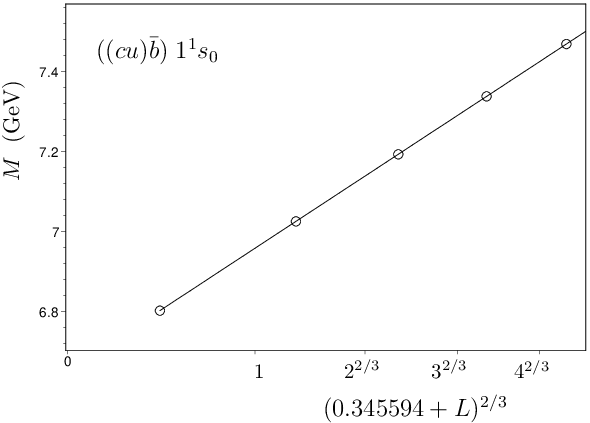}}
\subfigure[]{\label{subfigure:cfa}\includegraphics[scale=0.3]{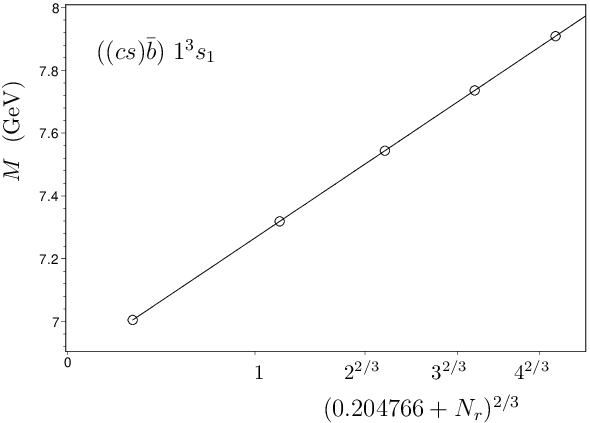}}
\subfigure[]{\label{subfigure:cfa}\includegraphics[scale=0.3]{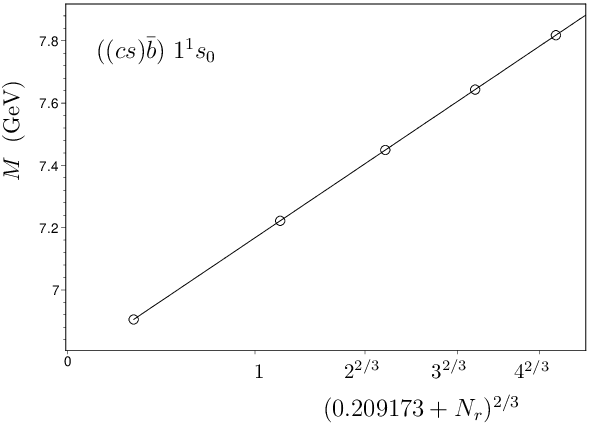}}
\subfigure[]{\label{subfigure:cfa}\includegraphics[scale=0.3]{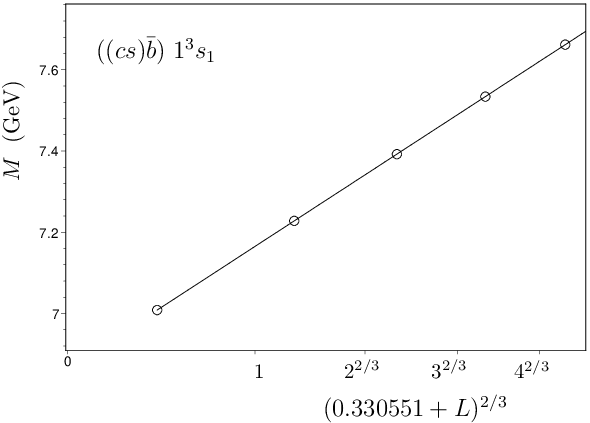}}
\subfigure[]{\label{subfigure:cfa}\includegraphics[scale=0.3]{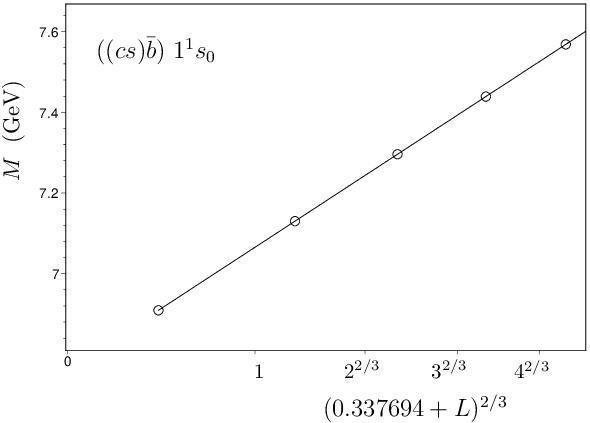}}
\subfigure[]{\label{subfigure:cfa}\includegraphics[scale=0.3]{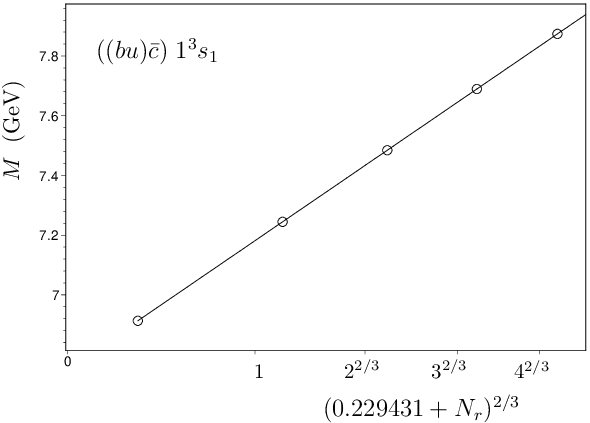}}
\subfigure[]{\label{subfigure:cfa}\includegraphics[scale=0.3]{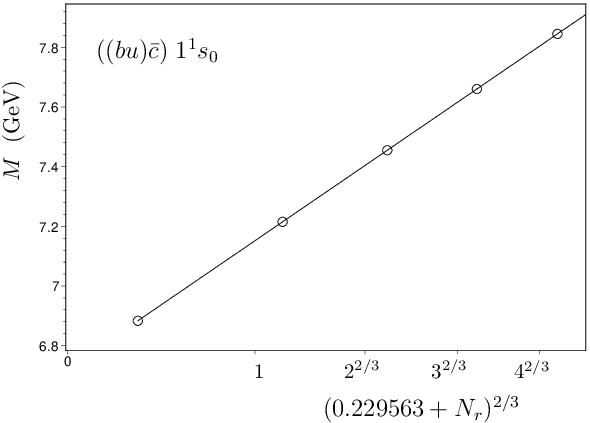}}
\subfigure[]{\label{subfigure:cfa}\includegraphics[scale=0.3]{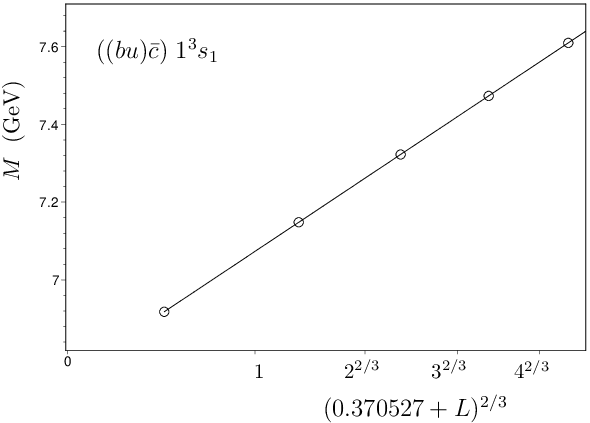}}
\subfigure[]{\label{subfigure:cfa}\includegraphics[scale=0.3]{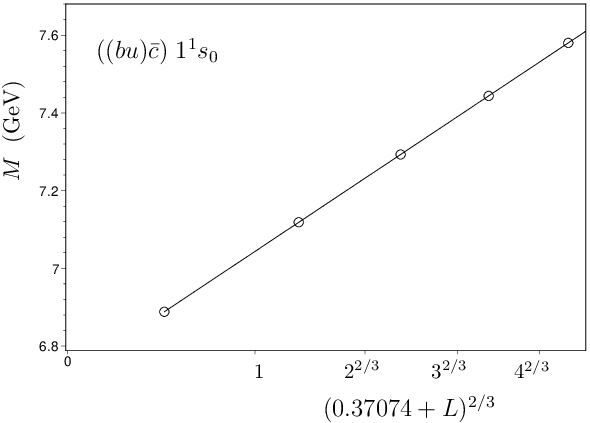}}
\subfigure[]{\label{subfigure:cfa}\includegraphics[scale=0.3]{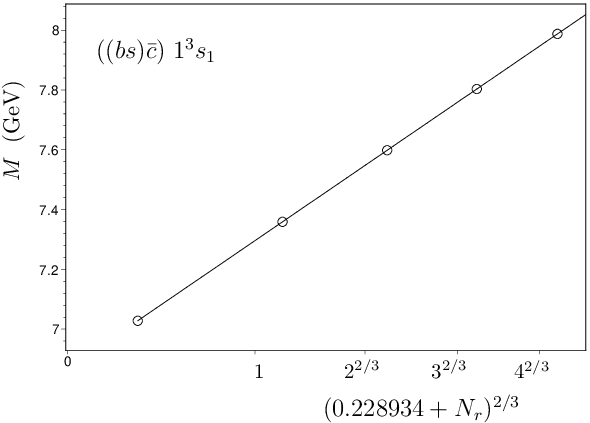}}
\subfigure[]{\label{subfigure:cfa}\includegraphics[scale=0.3]{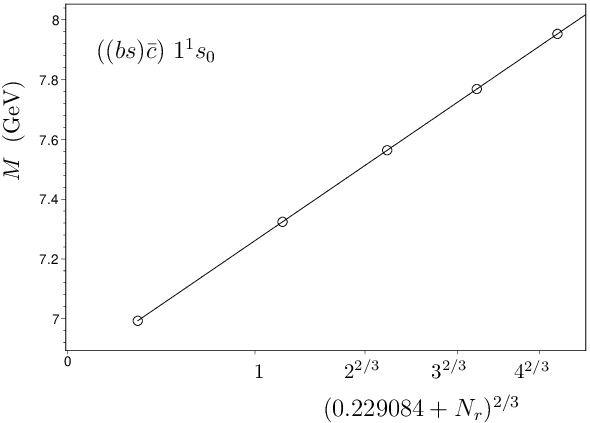}}
\subfigure[]{\label{subfigure:cfa}\includegraphics[scale=0.3]{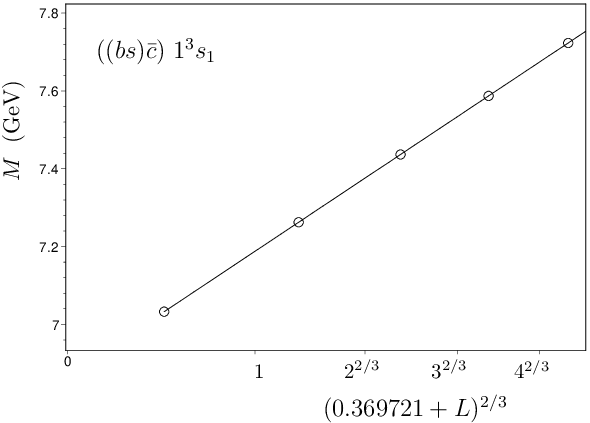}}
\subfigure[]{\label{subfigure:cfa}\includegraphics[scale=0.3]{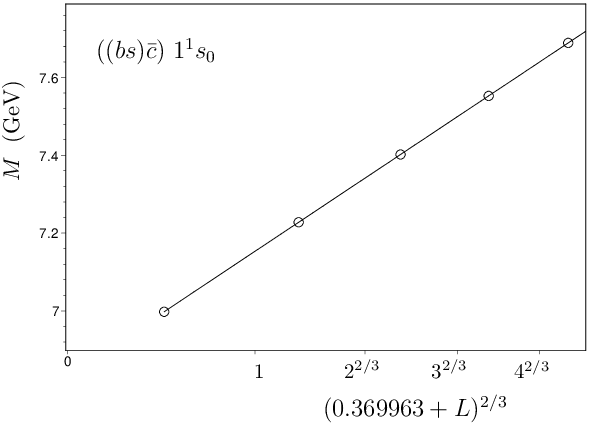}}
\subfigure[]{\label{subfigure:cfa}\includegraphics[scale=0.3]{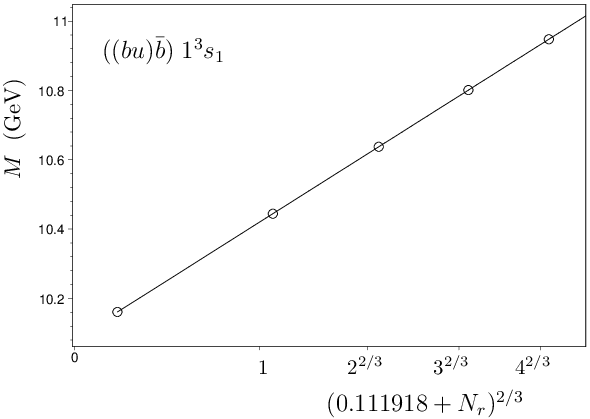}}
\subfigure[]{\label{subfigure:cfa}\includegraphics[scale=0.3]{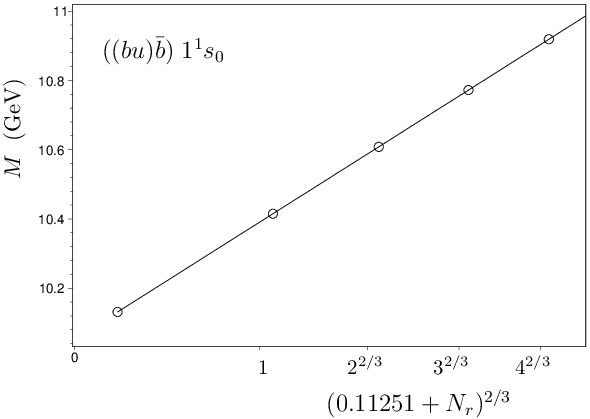}}
\subfigure[]{\label{subfigure:cfa}\includegraphics[scale=0.3]{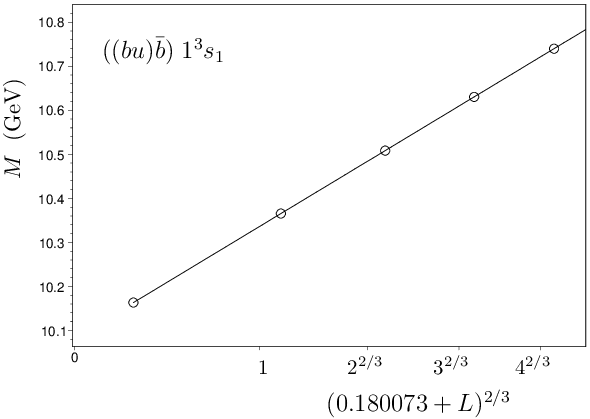}}
\subfigure[]{\label{subfigure:cfa}\includegraphics[scale=0.3]{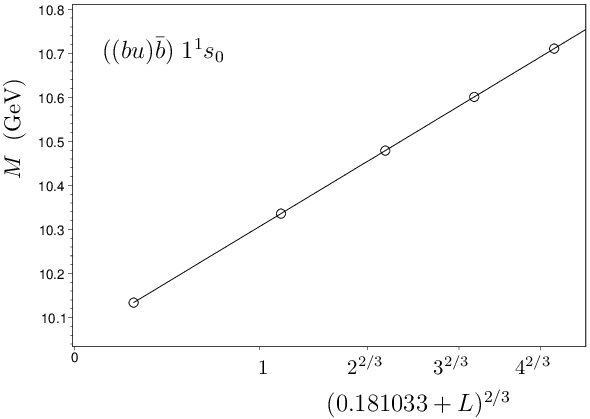}}
\subfigure[]{\label{subfigure:cfa}\includegraphics[scale=0.3]{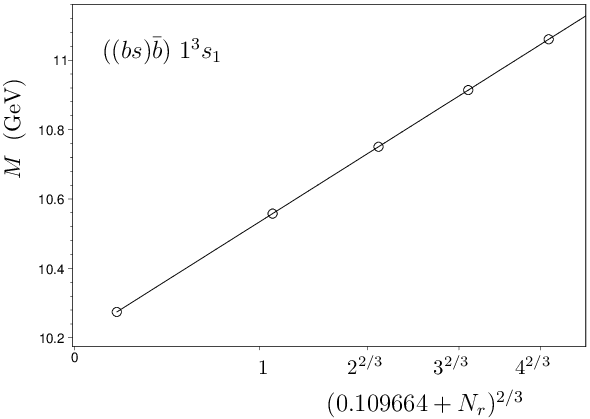}}
\subfigure[]{\label{subfigure:cfa}\includegraphics[scale=0.3]{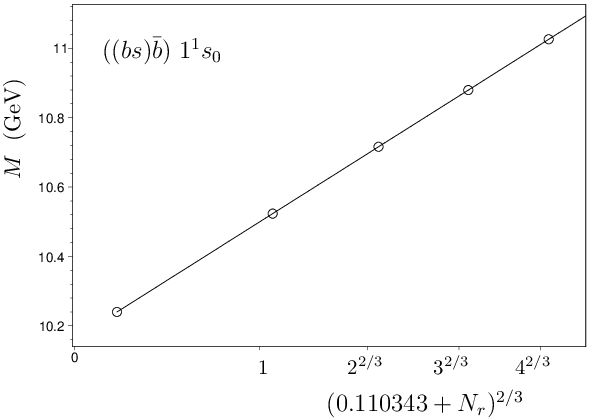}}
\subfigure[]{\label{subfigure:cfa}\includegraphics[scale=0.3]{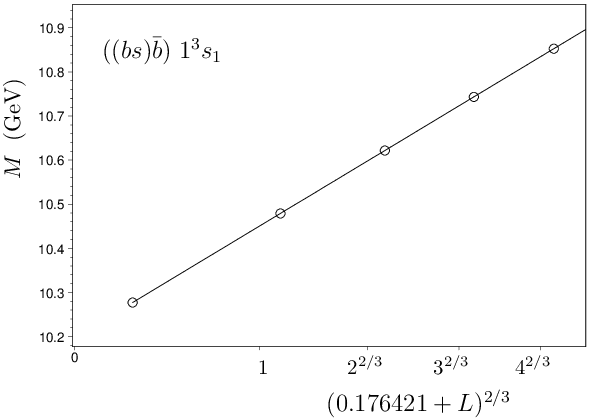}}
\subfigure[]{\label{subfigure:cfa}\includegraphics[scale=0.3]{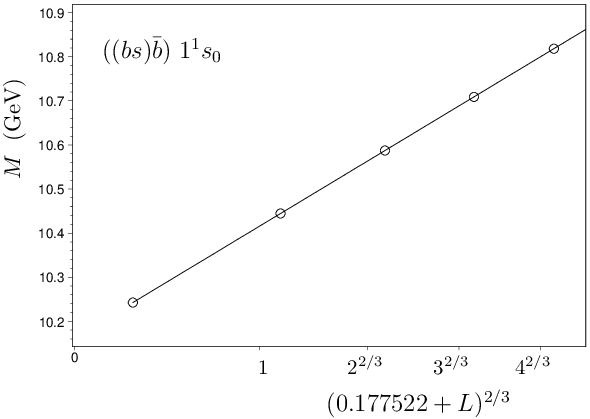}}
\caption{Radial and orbital $\lambda$-{\trs} for the {\dhts} $((cu)\bar{c})$, $((cu)\bar{b})$, $((cs)\bar{c})$, $((cs)\bar{b})$, $((bu)\bar{c})$, $((bu)\bar{b})$, $((bs)\bar{c})$, and $((bs)\bar{b})$. $1^1s_0$ and $1^3s_1$ denote the spin singlets and spin triplets of diquarks, respectively. Circles represent the predicted data and the black lines represent the fitted formulas. Data are listed in Tables \ref{tab:masslambdac} and the formulas are in Table \ref{tab:rtbehav}.}\label{fig:lambda}
\end{figure*}

When calculating the $\lambda$-mode radially and orbitally excited states, the $\rho$-mode state is taken as the radial ground state.
Using Eqs. (\ref{t2q}), (\ref{pa2qQ}), (\ref{fitcfxnr}) [or Eqs. (\ref{combrt}), (\ref{pa2qQ}), (\ref{fitcfxnr})], and parameters in Table \ref{tab:parmv}, the spectra of the $\lambda$-excited states of the {\dhts} $((cu)\bar{c})$, $((cu)\bar{b})$, $((cs)\bar{c})$, $((cs)\bar{b})$, $((bu)\bar{c})$, $((bu)\bar{b})$, $((bs)\bar{c})$, and $((bs)\bar{b})$ are calculated, see Table \ref{tab:masslambdac}. The $\lambda$-trajectories for the {\dhts} are shown in Fig. \ref{fig:lambda}.

The masses of the radially and orbitally $\lambda$-excited states $((bq)\bar{c})$ $(q=u,d,s)$ are greater than those of the corresponding $\lambda$-excited states $((cq)\bar{b})$ (see Tables \ref{tab:massrho} and \ref{tab:masslambdac}). This is because the binding energy between a heavy diquark and a heavy antiquark increase with the decreasing of reduced mass, see Eqs. (\ref{pa2qQ}) and (\ref{fitcfxnr}).
Accordingly, the following inequalities hold for the {\dhts}:
\begin{align}\label{masseq}
M\left(\lambda, ((bu)\bar{c})\right)>
M\left(\lambda, ((cu)\bar{b})\right),\nonumber\\
M\left(\lambda, ((bs)\bar{c})\right)>
M\left(\lambda, ((cs)\bar{b})\right),
\end{align}
where $M\left(\lambda, ((bu)\bar{c})\right)$ denotes the masses of the radially or orbitally $\lambda$-mode excited state of $((bu)\bar{c})$; the other notations in Eq. (\ref{masseq}) are similarly defined.

Unlike the $\rho$-{\trs} [see Eqs. (\ref{combrt}) and (\ref{pa2qQ})], the complete forms of the $\lambda$-{\trs} take simple forms, see Eq. (\ref{lambdform}). For the $\lambda$-{\trs}, the main parts are same as the complete forms because the diquark is taken as a constituent and the diquark's substructure is not considered when calculating the $\lambda$-trajectories. Therefore, the fitted formulas of the $\lambda$-{\trs} are same as the complete forms. From Table \ref{tab:rtbehav}, we can see that the $\lambda$-{\trs} follow the behavior $M{\sim}x_{\lambda}^{2/3}$, where $x_{\lambda}=L,\,N_r$.

\subsection{Discussions}

Triquarks are colored states and are not observable. The observable is the hadron, for example, the pentaquark. Analogous to mesons, which consist of one $3_c$ quark and one $\bar{3}_c$ antiqurk, a color-neutral pentaquark in the triquark-diquark picture can be constructed from a triquark in color $3_c$ and a diquark in color {\cltb}. In the triquark-diquark picture, triquark and diquark are taken as constituents forming pentaquarks. By analogy with meson case, the color Casimir for interaction between a $3_c$ triquark and a $\bar{3}_c$ diqurk is $-4/3$. By employing the obtained triquark {\rt} relations along with the diquark {\rt} relations, Eqs. (\ref{t2q}) and (\ref{pa2qQ}) or Eqs. (\ref{combrt}) and (\ref{pa2qQ}), we crudely estimate the spin-averaged masses of the ground state of pentaquarks  $(\bar{c}(cu))(cu)$, $(\bar{b}(bu))(bu)$, and $(\bar{c}(cu))(bu)$. The obtained results are consistent with other theoretical prediction, see Table \ref{tab:masscomp}.

\begin{table}[!phtb]
\caption{Comparison of theoretical predictions for the spin-averaged masses of the ground state of pentaquarks (in {\gev}). $[Qu]$ denotes spin-0 diquark $(Qu)$ while $\{Qu\}$ denotes spin-1 diquark $(Qu)$. ${\ast}$ denotes that the discussed pentaquarks are $QQQu\bar{u}$. $\dag$ denotes the state $(\bar{c}[bu])\{cu\}$.}  \label{tab:masscomp}
\centering
\begin{tabular*}{0.48\textwidth}{@{\extracolsep{\fill}}cccc@{}}
\hline\hline
\addlinespace[2pt]
    & $(\bar{c}(cu))(cu)$    & $(\bar{b}(bu))(bu)$   &  $(\bar{c}\{cu\})[bu]$ \\
\hline\\[-2.5mm]
 $(\bar{Q}\{Qu\})[Qu]$           &5.502    &15.231    &8.811   \\
 $(\bar{Q}[Qu])\{Qu\}$           &5.500    &15.230    &8.835$^{\dag}$   \\
 \cite{Ongodo:2025dad}           &6.643    &16.090    &   \\
 \cite{MoosaviNejad:2020hhd}     &6.017    &15.99     &9.557   \\
 \cite{Sharma:2024pfi}           &5.824    &16.220    &   \\
 \cite{Li:2018vhp}$^{\ast}$      &5.964    &15.917   &9.327   \\
 \cite{Abu-Shady:2024jmg}        &6.377    &16.03    &   \\
\hline\hline
\end{tabular*}
\end{table}

According to discussions in \ref{subsec:prelim}, Eqs. (\ref{t2q}) and (\ref{pa2qQ}) or Eqs. (\ref{combrt}) and (\ref{pa2qQ}), for the {\dhts} $((Qq)\bar{Q}')$, there are two series of spectra: the spectra of the $\lambda$-excited states and the spectra of the $\rho$-excited states. Correspondingly, there are two series of {\rts}: the $\lambda$-trajectories and the $\rho$-trajectories.
For the triply heavy triquarks \cite{Song:2024bkj}, the $\lambda$-trajectories
and the $\rho$-trajectories have the same behaviors, $M{\sim}x^{2/3}_{\lambda}$ and $M{\sim}x^{2/3}_{\rho}$, respectively. Different from the {\rts} for the triply heavy triquarks, for the {\dhts} $((Qq)\bar{Q}')$, the $\lambda$-trajectories follow the behavior $M{\sim}x_{\lambda}^{2/3}$ ($x_{\lambda}=L,\,N_r$) while the $\rho$-{\trs} exhibit the behavior $M{\sim}\sqrt{x_{\rho}}$ ($x_{\rho}=n_{r},\;l$).

Similar to the meson {\rts}, the $\lambda$-{\trs} for the {\dhts} $((Qq)\bar{Q}')$ take simple forms. However, for the $\rho$-{\trs}, it is noteworthy that the main parts of the complete forms and the fitted formulas share the same functional form, $M=m_R+\beta\sqrt{x+c_0}$. However, the values of the parameters $m_R$ and $\beta$ differ considerably.
For example, for the main part of the complete form of the radial $\rho$-{\tr} for the states $|n^3s_1, 1S\rangle$ of triquark $((cu)\bar{c})$, $m_R=3.13$ and $\beta=0.751988$ (see Eqs. (\ref{fulla}) and (\ref{fullb})); these values are readily computed using Eqs. (\ref{t2q}), (\ref{pa2qQ}), and (\ref{fitcfxnr}), or Eqs. (\ref{combrt}), (\ref{pa2qQ}), and (\ref{fitcfxnr}). By contrast, for the fitted $\rho$-{\tr}, $m_R=3.3257$ and $\beta=0.741064$ (see Eq. (\ref{fra})). As discussed above and in Sec. \ref{subsec:rho}, the $\rho$-{\trs} for {\dhts} cannot be obtained by merely mimicking the meson {\rts}; instead, they should be constructed based on the actual structure and substructure of the triquark.
Otherwise, the $\rho$-trajectories must rely solely on fitting existing theoretical or future experimental data.

In general, when the number of quarks exceeds 3, there will be different configurations and pictures. For example, there are two configurations for baryon $\Omega_{ccb}$ in the diquark picture \cite{Cakir:2026fzd}: $b+(cc)$, $c+(bc)$.
Tetraquarks can be studied in diquark-antidiquark picture \cite{Faustov:2021hjs}, quark-triquark picture \cite{Yasui:2007dv}, constituent picture \cite{Lu:2020rog}, and other pictures. Similar to baryon case \cite{Cakir:2026fzd}, in the quark-triquark picture, there are different configurations for tetraquarks: $q_1(q_2(\bar{q}_3\bar{q}_4))$, $q_2(q_1(\bar{q}_3\bar{q}_4))$, $((q_1q_2)\bar{q}_3)\bar{q}_4$, and $((q_1q_2)\bar{q}_4)\bar{q}_3$. In Ref. \cite{Lu:2020rog}, both color triplet and color sextet are considered, and quarks belonging to separate clusters are permitted to couple with each other; as a result, the color factor value $\langle \textbf{F}_{i}\cdot \textbf{F}_{j}\rangle$ is no longer unique. In the present work, only color triplet and antitriplet are considered. Furthermore, diquarks and triquarks are taken as constituents. Quarks inside diquark or triquark clusters do not couple to quarks residing in other clusters \cite{Faustov:2021hjs}. Consequently, the color factor of a triquark is simple and has the single value, as summarized in Table \ref{tab:casimir}.

Baryons $QQ'q$ have two configurations in the diquark picture: $(Qq){Q}'$ and $(QQ'){q}$. Correspondingly, there are two types of triquarks: $(Qq)\bar{Q'}$ and $(QQ')\bar{q}$. The {\rts} for baryon $(QQ'){q}$ differ from those for triquark $(QQ')\bar{q}$, and the difference arises from different values of color-Casimir. For {\dhts} $((Qq)\bar{Q}')$, the {\rt} relation is given in Eq. (\ref{combrt}). The baryon counterpart corresponding to the triquark $((Qq)\bar{Q}')$ is $(Qq){Q}'$. The Casimir is $-4/3$ for diquark-quark interaction, whereas it is $-2/3$ for diquark-antiquark interaction. Therefor, the {\rt} relation for the configuration $(Qq){Q}'$ ($Q{\ne}Q'$) reads
\begin{align}
M=&m_{Q}+m_{q}+m_{Q'}+\frac{3}{2}C\nonumber\\
&+\beta'_{x_{\lambda}}(x_{\lambda}+c_{0x_{\lambda}})^{2/3}
+\beta_{x_{\rho}}\sqrt{x_{\rho}+c_{0x_{\rho}}},
\end{align}
where
\bea
\beta'_{L}=\frac{3}{2}\left(\frac{\sigma^2}{\mu_{\lambda}}\right)^{1/3}c_{fL},\; \beta'_{N_r}=\frac{(3\pi)^{2/3}}{2}\left(\frac{\sigma^2}{\mu_{\lambda}}\right)^{1/3}c_{fN_r}.\nonumber
\eea
For configuration $(Qq){Q'}$ ($Q=Q'$), mixing between $\lambda$-mode and $\rho$-mode arises owing to the identical quarks $Q=Q'$. In contrast, no mixing between the $\lambda$-mode and $\rho$-mode occurs for the triquark $(Qq)\bar{Q'}$ even when $Q=Q'$ since heavy quark $Q$, light quark $q$, and heavy antiquark $\bar{Q}$ are distinguishable.

\section{Conclusions}\label{sec:conc}
In this work, we attempt to apply the Regge trajectory approach to the {\dhts} $((Qq)\bar{Q}^{\prime})$ $(Q,\,Q'=b,\,c;\,q=u,d,s)$. We propose the Regge trajectory relations for the {\dhts}, and then employ them to crudely estimate the spectra of the {\dhts} $((cu)\bar{c})$, $((cu)\bar{b})$, $((cs)\bar{c})$, $((cs)\bar{b})$, $((bu)\bar{c})$, $((bu)\bar{b})$, $((bs)\bar{c})$, and $((bs)\bar{b})$.

The $\lambda$- and $\rho$-trajectories for the {\dhts} are investigated. For the {\dhts} $((Qq)\bar{Q}^{\prime})$, the $\lambda$-trajectories follow the behavior $M{\sim}x_{\lambda}^{2/3}$ ($x_{\lambda}=L,\,N_r$), while the $\rho$-trajectories follow the behavior $M{\sim}\sqrt{x_{\rho}}$ ($x_{\rho}=l,\,n_r$).
The triquark Regge trajectory is a new and very simple approach for estimating the spectra of triquarks. It also provides a simple method to investigate easily the $\rho$-mode and $\sigma$-mode excitations of pentaquarks and hexaquarks in the triquark picture.

Triquarks are colored states and are not observable. In the diquark-triquark model, the colorless pentaquarks are composed of one triquark and one diquark, and they can be observed. We crudely estimate the spin-averaged masses of the ground states of the pentaquarks $(\bar{c}(cu))(cu)$, $(\bar{b}(bu))(bu)$, and $(\bar{c}(cu))(bu)$ by employing the obtained triquark {\rts}. The estimated results are consistent with other theoretical predictions.

\hspace{1cm}

\noindent {\bf CRediT authorship contribution statement}

\noindent {\bf Xin-Ru Liu}: Investigation, Validation, Data curation, Writing -- review \& editing. {\bf Qi Liu}: Methodology, Data curation, Writing -- original draft. {\bf He Song}: Conceptualization, Writing -- original draft. {\bf Jiao-Kai Chen}: Supervision, Methodology, Conceptualization.

\hspace{1cm}

\noindent {\bf Declaration of competing interest}

\noindent The authors declare that they have no known competing financial interests or personal relationships that could have appeared to influence the work reported in this paper.

\hspace{1cm}

\noindent {\bf Acknowledgments}

\noindent We are very grateful to the anonymous referees for the valuable comments and suggestions.

\hspace{1cm}

\noindent {\bf Data availability}

\noindent Data will be made available on request.

\appendix

\section{States of triquarks}
The states of triquarks in the diquark picture are listed in Table \ref{tab:bqstates}.

\begin{table*}[!phtb]
\caption{The states of triquarks in $3_c$ composed of one diquark in $\cltba$ and one antiquark in $\cltba$ \cite{Song:2024bkj}. The notation is explained in \ref{subsec:prelim}. Here, $q$, $\qp$ and $\qpp$ represent both the light quarks and the heavy quarks.}  \label{tab:bqstates}
\centering
\begin{tabular*}{\textwidth}{@{\extracolsep{\fill}}ccccc@{}}
\hline\hline
$J^P$ & $(L,l)$  &  Configuration \\
\hline
$\frac{1}{2}^+$ & $(0,0)$ & $\left({\qqs}^{\cltba}_{n^1s_0}{\qpp}\right)^{3_c}_{N^2S_{1/2}}$,\;
$\left({\qqb}^{\cltba}_{n^3s_1}{\qpp}\right)^{3_c}_{N^2S_{1/2}}$,\\
& $(1,1)$ &
$\left({\qqb}^{\cltba}_{n^1p_1}{\qpp}\right)^{3_c}_{N^2P_{1/2}}$,\;
$\left({\qqb}^{\cltba}_{n^1p_1}{\qpp}\right)^{3_c}_{N^4P_{1/2}}$,\;
$\left({\qqs}^{\cltba}_{n^3p_0}{\qpp}\right)^{3_c}_{N^2P_{1/2}}$,\;
$\left({\qqs}^{\cltba}_{n^3p_1}{\qpp}\right)^{3_c}_{N^2P_{1/2}}$,\\
& &
$\left({\qqs}^{\cltba}_{n^3p_1}{\qpp}\right)^{3_c}_{N^4P_{1/2}}$,\;
$\left({\qqs}^{\cltba}_{n^3p_2}{\qpp}\right)^{3_c}_{N^4P_{1/2}}$\\
&$\cdots$ &$\cdots$ \\
$\frac{1}{2}^-$ & $(1,0)$ &
$\left({\qqs}^{\cltba}_{n^1s_0}{\qpp}\right)^{3_c}_{N^2P_{1/2}}$,\;
$\left({\qqb}^{\cltba}_{n^3s_1}{\qpp}\right)^{3_c}_{N^2P_{1/2}}$,\;
$\left({\qqb}^{\cltba}_{n^3s_1}{\qpp}\right)^{3_c}_{N^4P_{1/2}}$,\\
& $(0,1)$ &
$\left({\qqb}^{\cltba}_{n^1p_1}{\qpp}\right)^{3_c}_{N^2S_{1/2}}$,\;
$\left({\qqs}^{\cltba}_{n^3p_0}{\qpp}\right)^{3_c}_{N^2S_{1/2}}$,\;
$\left({\qqs}^{\cltba}_{n^3p_1}{\qpp}\right)^{3_c}_{N^2S_{1/2}}$,\\
&$\cdots$ &$\cdots$ \\
$\frac{3}{2}^+$ & $(0,0)$ &
$\left({\qqb}^{\cltba}_{n^3s_1}{\qpp}\right)^{3_c}_{N^4S_{3/2}}$,\\
& $(1,1)$ &
$\left({\qqb}^{\cltba}_{n^1p_1}{\qpp}\right)^{3_c}_{N^2P_{3/2}}$,\;
$\left({\qqb}^{\cltba}_{n^1p_1}{\qpp}\right)^{3_c}_{N^4P_{3/2}}$,\;
$\left({\qqs}^{\cltba}_{n^3p_0}{\qpp}\right)^{3_c}_{N^2P_{3/2}}$,\;
$\left({\qqs}^{\cltba}_{n^3p_1}{\qpp}\right)^{3_c}_{N^2P_{3/2}}$,\;\\
& &
$\left({\qqs}^{\cltba}_{n^3p_1}{\qpp}\right)^{3_c}_{N^4P_{3/2}}$,\;
$\left({\qqs}^{\cltba}_{n^3p_2}{\qpp}\right)^{3_c}_{N^4P_{3/2}}$,\;
$\left({\qqs}^{\cltba}_{n^3p_2}{\qpp}\right)^{3_c}_{N^6P_{3/2}}$,\\
&$\cdots$ &$\cdots$ \\
$\frac{3}{2}^-$ & $(1,0)$ &
$\left({\qqs}^{\cltba}_{n^1s_0}{\qpp}\right)^{3_c}_{N^2P_{3/2}}$,\;
$\left({\qqb}^{\cltba}_{n^3s_1}{\qpp}\right)^{3_c}_{N^2P_{3/2}}$,\;
$\left({\qqb}^{\cltba}_{n^3s_1}{\qpp}\right)^{3_c}_{N^4P_{3/2}}$,\\
& $(0,1)$ &
$\left({\qqb}^{\cltba}_{n^1p_1}{\qpp}\right)^{3_c}_{N^4S_{3/2}}$,\;
$\left({\qqs}^{\cltba}_{n^3p_1}{\qpp}\right)^{3_c}_{N^4S_{3/2}}$,\;
$\left({\qqs}^{\cltba}_{n^3p_2}{\qpp}\right)^{3_c}_{N^4S_{3/2}}$\\
&$\cdots$ &$\cdots$ \\
\hline\hline
\end{tabular*}
\end{table*}

\section{List of the {\rt} relations}\label{sec:appa}
The concrete forms of the {\rts} for {\dhts} are not as simple as those for mesons. Although Eqs. (\ref{t2q}) and (\ref{pa2qQ}) or Eqs. (\ref{combrt}) and (\ref{pa2qQ}) are compact, their final forms will be rather tedious due to triquark substructures and the mass dependence of the $\lambda$-trajectory slope.

From Eqs. (\ref{t2q}) and (\ref{pa2qQ}), or from Eqs. (\ref{pa2qQ}) and (\ref{combrt}), we can easily obtain the complete forms of the {\rts} for the {\dhts} $((Qq)\bar{Q}^{\prime})$. The resulting expressions are rather long and tedious. As an example, we list the radial $\rho$-{\tr} for the states $|n^3s_1, 1S\rangle$ of triquark $((cu)\bar{c})$, which reads
\begin{widetext}
\begin{align}\label{fulla}
M=&3.13+0.751988\sqrt{0.17+n_r}+ 0.38714\left(1.73+0.751988\sqrt{0.17+n_r}\right)^{-1/3}\left(3.28+0.751988\sqrt{0.17+n_r}\right)^{1/3} \nonumber\\
&\Big(0.334-0.13485\left(1.73+0.751988\sqrt{0.17+n_r}\right)
\left(3.28+0.751988\sqrt{0.17+n_r}\right)^{-1}\Big)^{2/3}\nonumber\\
 &\Big(1.008+0.0124\left(1.73+0.751988\sqrt{0.17+n_r}\right)
\left(3.28+0.751988\sqrt{0.17+n_r}\right)^{-1}\Big).
\end{align}
The corresponding main part of the full form (\ref{fulla}) is
\bea\label{fullb}
M=3.13+0.751988\sqrt{0.17+n_r}.
\eea
The fitted formula corresponding to Eq. (\ref{fulla}) is
\bea\label{fra}
M=3.3257+0.741064 \sqrt{0.17+ n_{r}}.
\eea
\end{widetext}
The $\rho$-excited masses for the states $|n^3s_1, 1S\rangle$ of triquark $((cu)\bar{c})$ can be calculated using Eq. (\ref{fulla}); the corresponding results are listed in Table \ref{tab:massrho}. By linearly fitting the calculated masses in the $(M,\sqrt{c0+ x})$ plane, we obtain the fitted formulas in Eq. (\ref{fra}). $m_R$ and slope of the fitted {\rt} [Eq. (\ref{fra})] are different from those of the main parts of the complete forms of the {\rt} [Eq. (\ref{fulla}) or (\ref{fullb})].

The explicit forms of the {\rts} calculated from Eqs. (\ref{t2q}) and (\ref{pa2qQ}), or from Eqs. (\ref{combrt}) and (\ref{pa2qQ}) are often rather tedious. Here, we present only the fitted formulas obtained by fitting the calculated results (see Table \ref{tab:rtbehav}).

\begin{table*}[!phtb]
\caption{Fitted formulas of the $\rho$- and $\lambda$-{\trs} for the {\dhts} $((Qq)\bar{Q})$. The fitted formulas are obtained by fitting the numerical results calculated using Eqs. (\ref{t2q}) and (\ref{pa2qQ}) or Eqs. (\ref{combrt}) and (\ref{pa2qQ}). The notation in Eq. (\ref{tetnot}) is rewritten as $|n^{2s_d+1}l_{j_d},N^{2s_t+1}L_J\rangle$. And $|n^{2s_d+1}l_{j_d},NL\rangle$ denotes the spin-averaged states. $n^1s_0$, $n^3s_1$: radial {\rts} for spin-0 and spin-1 diquarks; $1^1l_{l}$, $1^3l_{l+1}$: orbital {\rts} for spin-0 and spin-1 diquarks.}\label{tab:rtbehav}
\centering
\begin{tabular*}{1.0\textwidth}{@{\extracolsep{\fill}}ccccc@{}}
\hline\hline
  &   \multicolumn{2}{c}{$\rho$-{\trs}} &   \multicolumn{2}{c}{$\lambda$-{\trs}}    \\
      & $|n^{2s_d+1}l_{j_d},NL\rangle$ &  {\trs}  & $|n^{2s_d+1}l_{j_d},NL\rangle$ &  {\trs}  \\
\hline
$((cu)\bar{c})$
    & $|n^3s_1,1S\rangle$ & $M=3.3257+0.741064 \sqrt{0.17+ n_{r}}$
    & $|1^3s_1,NS\rangle$ &  $M=3.44005+0.474432 (0.257371+N_r)^{2/3}$ \\
    & $|n^1s_0,1S\rangle$ & $M=3.32663+0.740458 \sqrt{0.065+ n_{r}}$
    & $|1^1s_0,NS\rangle$ &  $M=3.32172+0.478514 (0.259356+N_r)^{2/3}$ \\
    & $|1^3l_{l+1},1S\rangle$ & $M=3.3265+0.612951 \sqrt{0.19+l}$
    & $|1^3s_1,1L\rangle$ &  $M=3.44005+0.354318 (0.415809+L)^{2/3}$ \\
    & $|1^1l_l,1S\rangle$ & $M=3.32707+0.612577 \sqrt{0.095+l}$
    & $|1^1s_0,1L\rangle$ &  $M=3.32172+0.357336 (0.419025+L)^{2/3}$ \\
$((cs)\bar{c})$
    & $|n^3s_1,1S\rangle$ & $M=3.49241+0.753776 \sqrt{0.095+n_r}$
    & $|1^3s_1,NS\rangle$ &  $M=3.53549+0.471425 (0.255864+N_r)^{2/3}$ \\
    & $|n^1s_0,1S\rangle$ & $M=3.49313+0.753311 \sqrt{0.03+n_r}$
    & $|1^1s_0,NS\rangle$ &  $M=3.43233+0.474686 (0.257497+N_r)^{2/3}$ \\
    & $|1^3l_{l+1},1S\rangle$ & $M=3.49297+0.600131 \sqrt{0.135+l}$
    & $|1^3s_1,1L\rangle$ &  $M=3.53549+0.352094 (0.413365+L)^{2/3}$ \\
    & $|1^1l_l,1S\rangle$ & $M=3.49346+0.599809 \sqrt{0.055+l}$
    & $|1^1s_0,1L\rangle$ &  $M=3.43233+0.354505 (0.416012+L)^{2/3}$ \\
$((cu)\bar{b})$
    & $|n^3s_1,1S\rangle$ & $M=6.60968+0.731915 \sqrt{0.17+n_r}$
    & $|1^3s_1,NS\rangle$ &  $M=6.77005+0.404625 (0.208838+N_r)^{2/3}$ \\
    & $|n^1s_0,1S\rangle$ & $M=6.6109+0.731126 \sqrt{0.065+n_r}$
    & $|1^1s_0,NS\rangle$ &  $M=6.65172+0.410206 (0.214047+N_r)^{2/3}$ \\
    & $|1^3l_{l+1},1S\rangle$ & $M=6.61076+0.605101 \sqrt{0.19+l}$
    & $|1^3s_1,1L\rangle$ &  $M=6.77005+0.302797 (0.337152+L)^{2/3}$ \\
    & $|1^1l_l,1S\rangle$ & $M=6.61149+0.604619 \sqrt{0.095+l}$
    & $|1^1s_0,1L\rangle$ &  $M=6.65172+0.306907 (0.345594+L)^{2/3}$ \\
$((cs)\bar{b})$
    & $|n^3s_1,1S\rangle$ & $M=6.77405+0.744778 \sqrt{0.095+n_r}$
    & $|1^3s_1,NS\rangle$ &  $M=6.86549+0.400476 (0.204766+N_r)^{2/3}$ \\
    & $|n^1s_0,1S\rangle$ & $M=6.77499+0.744164 \sqrt{0.03+n_r}$
    & $|1^1s_0,NS\rangle$ &  $M=6.76233+0.404974 (0.209173+N_r)^{2/3}$ \\
    & $|1^3l_{l+1},1S\rangle$ & $M=6.77482+0.592699 \sqrt{0.135+l}$
    & $|1^3s_1,1L\rangle$ &  $M=6.86549+0.299742 (0.330551+L)^{2/3}$ \\
    & $|1^1l_l,1S\rangle$ & $M=6.77546+0.592278 \sqrt{0.055+l}$
    & $|1^1s_0,1L\rangle$ &  $M=6.76233+0.303053 (0.337694+L)^{2/3}$ \\
$((bu)\bar{c})$
    & $|n^3s_1,1S\rangle$ & $M=6.62165+0.74034 \sqrt{0.155+n_r}$
    & $|1^3s_1,NS\rangle$ &  $M=6.75251+0.428814 (0.229431+N_r)^{2/3}$ \\
    & $|n^1s_0,1S\rangle$ & $M=6.62168+0.740325 \sqrt{0.125+n_r}$
    & $|1^1s_0,NS\rangle$ &  $M=6.72268+0.428989 (0.229563+N_r)^{2/3}$ \\
    & $|1^3l_{l+1},1S\rangle$ & $M=6.62175+0.576854 \sqrt{0.22+l}$
    & $|1^3s_1,1L\rangle$ &  $M=6.75251+0.320623 (0.370527+L)^{2/3}$ \\
    & $|1^1l_l,1S\rangle$ & $M=6.62177+0.576842 \sqrt{0.18+l}$
    & $|1^1s_0,1L\rangle$ &  $M=6.72268+0.320752 (0.37074+L)^{2/3}$ \\
$((bs)\bar{c})$
    & $|n^3s_1,1S\rangle$ & $M= 6.79096+0.714196 \sqrt{0.11+n_r}$
    & $|1^3s_1,NS\rangle$ &  $M=6.86768+0.428156 (0.228934+N_r)^{2/3}$ \\
    & $|n^1s_0,1S\rangle$ & $M=6.79099+0.71418 \sqrt{0.08+n_r}$
    & $|1^1s_0,NS\rangle$ &  $M=6.8327+0.428353 (0.229084+N_r)^{2/3}$ \\
    & $|1^3l_{l+1},1S\rangle$ & $M=6.79104+0.549425 \sqrt{0.16+l}$
    & $|1^3s_1,1L\rangle$ &  $M=6.86768+0.320138 (0.369721+L)^{2/3}$ \\
    & $|1^1l_l,1S\rangle$ & $M=6.79106+0.549411 \sqrt{0.115+l}$
    & $|1^1s_0,1L\rangle$ &  $M=6.8327+0.320283 (0.369963+L)^{2/3}$ \\
$((bu)\bar{b})$
    & $|n^3s_1,1S\rangle$ & $M=9.8711+0.735432 \sqrt{0.155+n_r}$
    & $|1^3s_1,NS\rangle$ &  $M=10.0825+0.337146 (0.111918+N_r)^{2/3}$ \\
    & $|n^1s_0,1S\rangle$ & $M=9.87114+0.735404 \sqrt{0.125+n_r}$
    & $|1^1s_0,NS\rangle$ &  $M=10.0527+0.337429 (0.11251+N_r)^{2/3}$ \\
    & $|1^3l_{l+1},1S\rangle$ & $M=9.87128+0.572942 \sqrt{0.22+l}$
    & $|1^3s_1,1L\rangle$ &  $M=10.0825+0.253305 (0.180073+L)^{2/3}$ \\
    & $|1^1l_l,1S\rangle$ & $M=9.87131+0.572921 \sqrt{0.18+l}$
    & $|1^1s_0,1L\rangle$ &  $M=10.0527+0.253511 (0.181033+L)^{2/3}$ \\
$((bs)\bar{b})$
    & $|n^3s_1,1S\rangle$ & $M=10.0392+0.709533 \sqrt{0.11+n_r}$
    & $|1^3s_1,NS\rangle$ &  $M=10.1977+0.336081 (0.109664+N_r)^{2/3}$ \\
    & $|n^1s_0,1S\rangle$ & $M=10.0393+0.709502 \sqrt{0.08+n_r}$
    & $|1^1s_0,NS\rangle$ &  $M=10.1627+0.336401 (0.110343+N_r)^{2/3}$ \\
    & $|1^3l_{l+1},1S\rangle$ & $M=10.0394+0.545756 \sqrt{0.16+l}$
    & $|1^3s_1,1L\rangle$ &  $M=10.1977+0.252528 (0.176421+L)^{2/3}$ \\
    & $|1^1l_l,1S\rangle$ & $M=10.0394+0.545732 \sqrt{0.115+l}$
    & $|1^1s_0,1L\rangle$ &  $M=10.1627+0.252761 (0.177522+L)^{2/3}$ \\
\hline
\hline
\end{tabular*}
\end{table*}

\end{document}